\documentclass[sigconf]{acmart}

\usepackage{multirow}
\usepackage{algorithm}
\usepackage{algorithmic}
\usepackage{bm}
\usepackage{dsfont}
\usepackage{colortbl}
\usepackage{tabularx}

\definecolor{dodgerblue}{RGB}{78,144,255}

\AtBeginDocument{%
  }

\copyrightyear{2025} 
\acmYear{2025} 
\setcopyright{acmlicensed}
\acmConference[MM '25]{Proceedings of the 33rd ACM International Conference on Multimedia}{October 27--31, 2025}{Dublin, Ireland}
\acmBooktitle{Proceedings of the 33rd ACM International Conference on Multimedia (MM '25), October 27--31, 2025, Dublin, Ireland}
\acmISBN{979-8-4007-2035-2/2025/10}
\acmDOI{10.1145/3746027.3755798}
\begin{document}

\title{EEG-SCMM: Soft Contrastive Masked Modeling for Cross-Corpus EEG-Based Emotion Recognition}

\author{Qile Liu}
\affiliation{
  \institution{Shenzhen University}
  \city{Shenzhen}
  \country{China}
}
\email{liuqile2022@email.szu.edu.cn}

\author{Weishan Ye}
\affiliation{
  \institution{Shenzhen University}
  \city{Shenzhen}
  \country{China}
}
\email{2110246024@email.szu.edu.cn}

\author{Lingli Zhang}
\affiliation{
  \institution{Shenzhen University}
  \city{Shenzhen}
  \country{China}
}
\email{2023220036@email.szu.edu.cn}

\author{Zhen Liang}
\authornote{Corresponding author.}
\affiliation{
  \institution{Shenzhen University}
  \city{Shenzhen}
  \country{China}
}
\email{janezliang@szu.edu.cn}

\begin{abstract}
  Emotion recognition using electroencephalography (EEG) signals has attracted increasing attention in recent years. However, existing methods often lack generalization in cross-corpus settings, where a model trained on one dataset is directly applied to another without retraining, due to differences in data distribution and recording conditions. To tackle the challenge of cross-corpus EEG-based emotion recognition, we propose a novel framework termed \textbf{S}oft \textbf{C}ontrastive \textbf{M}asked \textbf{M}odeling (\textbf{SCMM}). Grounded in the theory of emotional continuity, SCMM integrates soft contrastive learning with a hybrid masking strategy to effectively capture emotion dynamics (refer to short-term continuity). Specifically, in the self-supervised learning stage, we propose a soft weighting mechanism that assigns similarity scores to sample pairs, enabling fine-grained modeling of emotional transitions and capturing the temporal continuity of human emotions. To further enhance representation learning, we design a similarity-aware aggregator that fuses complementary information from semantically related samples based on pairwise similarities, thereby improving feature expressiveness and reconstruction quality. This dual design contributes to a more discriminative and transferable representation, which is crucial for robust cross-corpus generalization. Extensive experiments on the SEED, SEED-IV, and DEAP datasets show that SCMM achieves state-of-the-art (SOTA) performance, outperforming the second-best method by an average accuracy of 4.26\% under both same-class and different-class cross-corpus settings. The source code is available at https://github.com/Kyler-RL/SCMM.
\end{abstract}

\ccsdesc[300]{Human-centered computing~HCI design and evaluation methods}
\ccsdesc[300]{Computing methodologies~Artificial intelligence}

\keywords{EEG; Emotion Recognition; Soft Contrastive Learning; Masked Modeling; Cross-Corpus}

\maketitle

\section{Introduction}
\label{sec:Introduction}{
Emotions are human attitudinal experiences and behavioral responses to objective things, closely related to an individual's health conditions and behavioral patterns \cite{wang2024dmmr}. Compared to speech \cite{singh2022systematic}, gestures \cite{noroozi2018survey}, and facial expressions \cite{canal2022survey}, electroencephalography (EEG) offers a more direct and objective measurement of human emotions by capturing brain activity across various scalp locations \cite{hu2019ten}. Therefore, researchers have increasingly emphasized EEG-based emotion recognition in recent years \cite{zhong2020eeg, zhao2021plug, zhang2022ganser, zhang2023unsupervised}, aiming to advance the development of affective brain-computer interfaces (aBCIs). However, three critical challenges remain to be addressed in current approaches.

\begin{figure*}
    \begin{center}
        \includegraphics[width=0.9\textwidth]{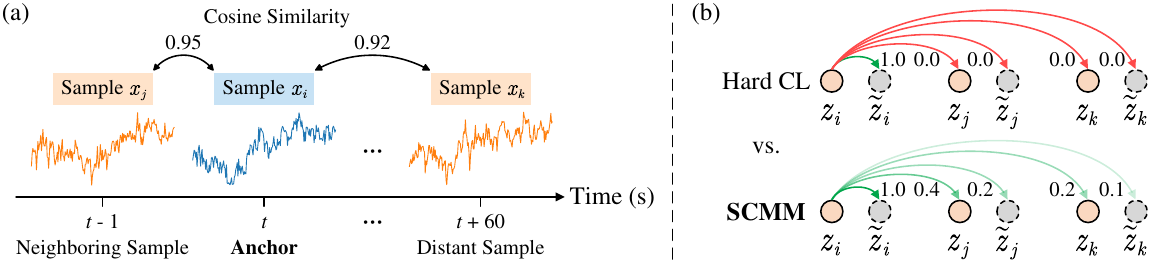}
    \end{center}
    \caption{(a) An illustration of emotional continuity. We take the sample $x_i$ at second $t$ within an EEG trial as the anchor, and calculate the cosine similarity between $x_i$ and its neighboring sample $x_j$, as well as the distant sample $x_k$. High cosine similarities indicate that human emotions remain relatively stable and similar over a certain period. (b) Hard CL vs. SCMM. Traditional hard CL assigns hard values (1 and 0) to positives and negatives when computing the contrastive loss. In contrast, our model generates soft assignments for different sample pairs, taking into account the emotional continuity.}
    \label{fig:Illustration}
    \Description{(a) An illustration of emotional continuity. (b) Hard CL vs. SCMM.}
\end{figure*}

(1) \textbf{Insufficient Generalization Capability.} Most existing EEG-based emotion recognition methods are typically designed for a single dataset, necessitating model retraining when the dataset changes. This requirement significantly limits the scalability and generalizability of the model, hindering its application on different datasets. To address this issue, the concept of \textbf{cross-corpus} has been proposed, which is designed to be generalized across multiple datasets. A cross-corpus model is trained on one dataset and can be directly applied to another without retraining from scratch. This concept, which originated in natural language processing \cite{schuller2010cross, zhang2011unsupervised}, has been extended to various domains in recent years \cite{rayatdoost2018cross, chien2020cross, ryumina2022search}. Although existing EEG-based emotion recognition methods, such as BiDANN \cite{li2018novel}, TANN \cite{li2021novel}, and PR-PL \cite{zhou2023pr}, have demonstrated superior performance in within-subject or cross-subject tasks within a single dataset, their effectiveness significantly degrades in cross-corpus scenarios, where the differences in data distribution across datasets far exceed the intra-dataset variability \cite{rayatdoost2018cross}.

(2) \textbf{Modeling Strategy Limitation.} Recently, researchers have explored domain adaptation techniques to address cross-corpus EEG-based emotion recognition \cite{he2022adversarial, zhou2025enhancing}. This is motivated by their efficacy to solve the problem of domain shift \cite{ganin2016domain}. Despite initial success, such approaches often require prior access to all labeled source data and unlabeled target data for model training. Considering the difficulties in collecting EEG signals as well as the time and expertise required to label them, the modeling strategy limitation of domain adaptation techniques poses a significant challenge to real-world aBCI applications.

(3) \textbf{Ignorance of Emotional Continuity}. Unlike domain adaptation techniques, contrastive learning (CL) achieves superior performance without relying on labeled data, and has shown great potential in various fields \cite{chen2020simple, radford2021learning, eldele2021time}. Current CL-based methods for EEG-based emotion recognition, such as CLISA \cite{shen2022contrastive} and JCFA \cite{liu2024joint}, consider an anchor and its augmented views as positive pairs, while treating all other samples as negatives. When computing the contrastive loss, the weights for positives and negatives are set to 1 and 0, respectively, as shown in Fig. \ref{fig:Illustration}(b) (Hard CL). However, psychological and neuroscientific studies suggest that emotion analysis using brain signals should account for dynamic changes \cite{davidson1998affective, houben2015relation}. Specifically, emotions exhibit significant "short-term continuity" characteristics, meaning that human emotions are relatively stable over certain periods, with sudden changes being rare. As illustrated in Fig. \ref{fig:Illustration}(a), a high cosine similarity is maintained between an anchor sample $x_i$ and its neighboring sample $x_j$, and even a distant sample $x_k$ separated by extended periods (e.g., 60 seconds). Given this nature of emotions, we propose that the definition of positive pairs in CL-based EEG emotion analysis should extend beyond just the anchor and its augmented views. Instead, it should include a broader range of similar samples, especially those that are temporally proximal, as shown in Fig. \ref{fig:Illustration}(b) (SCMM). In contrast, existing methods following the traditional CL paradigm \cite{chen2020simple} may incorrectly pull apart similar but not identical samples, thus failing to capture the emotional continuity inherent in EEG signals.

To tackle the aforementioned three critical issues, we propose a novel \textbf{S}oft \textbf{C}ontrastive \textbf{M}asked \textbf{M}odeling (\textbf{SCMM}) framework for cross-corpus EEG-based emotion recognition. Unlike traditional hard CL shown in Fig. \ref{fig:Illustration}(b), SCMM considers emotional continuity and incorporates soft assignments of sample pairs. This approach enables the model to identify the fine-grained relationships between different samples in a self-supervised manner, thereby enhancing the generalizability of EEG representations. Comprehensive experiments on three well-recognized datasets show that SCMM consistently achieves state-of-the-art (SOTA) performance, demonstrating its superior capability and stability. In summary, the main contributions of SCMM are outlined as follows:

\begin{itemize}
    \item We propose a novel SCMM framework to address three key challenges (insufficient generalization capability, modeling strategy limitation, and ignorance of emotional continuity) in cross-corpus EEG-based emotion recognition.
    \item Inspired by the nature of emotions, we introduce a soft weighting mechanism that assigns similarity scores to sample pairs to capture the similarity relationships between different samples. As a result, better feature representations of EEG signals are learned in a self-supervised manner.
    \item We develop a new hybrid masking strategy to generate diverse masked samples by considering both channel and feature relationships, which is essential for enhancing contrastive learning. In addition, we introduce a similarity-aware aggregator to fuse complementary information from semantically related samples, enabling fine-grained feature learning and improving the model's overall capability.
    \item We conduct extensive experiments on three well-known datasets (SEED, SEED-IV, and DEAP), demonstrating that SCMM achieves SOTA performance against 10 baselines, with an average accuracy improvement of 4.26\% under both same-class and different-class cross-corpus settings.
\end{itemize}
}

\section{Related Work}
\label{sec:Related Work}{
\subsection{EEG-Based Emotion Recognition}
\label{subsec:EEG-Based Emotion Recognition}{
Current approaches for EEG-based emotion recognition mainly rely on two types of experimental protocols: (1) subject-dependent and (2) subject-independent.

(1) The subject-dependent protocol trains and tests models using EEG data from the same subject within a single dataset. For example, Duan \textit{et al.} \cite{duan2013differential} extracted various emotion-related features from EEG signals and used support vector machine (SVM) and k-nearest neighbors (KNN) for emotion recognition. Similarly, Alsolamy \textit{et al.} \cite{alsolamy2016emotion} inputted power spectral density (PSD) features into an SVM classifier to predict emotions while listening to the Quran. In terms of deep learning models, Zheng \textit{et al.} \cite{zheng2015investigating} trained a deep belief network (DBN) using differential entropy (DE) features extracted from multi-channel EEG signals for subject-dependent emotion classification. Additionally, Song \textit{et al.} \cite{song2018eeg} proposed a dynamical graph convolutional network (DGCNN) that dynamically learns the intrinsic relationship between different EEG channels to enhance the model's discriminative ability. In general, the subject-dependent protocol tends to achieve superior performance due to its potential to introduce information leakage. However, this protocol fails to take into account the significant individual differences of EEG signals, thus limiting its practical applications. 

(2) The subject-independent protocol trains and tests models using EEG data from different subjects within a single dataset. Since transfer learning has demonstrated its potential in addressing the problem of domain shift, a series of methods have adopted it for subject-independent EEG-based emotion recognition. For example, Li \textit{et al.} \cite{li2018novel} proposed a bi-hemispheres domain adversarial neural network (BiDANN) that considers distribution shift between training and testing data and cerebral hemispheres. Following this, a novel transferable attention neural network (TANN) \cite{li2021novel} was introduced to learn the emotional discriminative information of EEG signals. To simultaneously adapt the marginal distribution and the conditional distribution, Li \textit{et al.} \cite{li2019domain} proposed a joint distribution adaptation network (JDA) for subject-independent EEG-based emotion recognition. Similarly, Chen \textit{et al.} \cite{chen2021ms} introduced a multi-source marginal distribution adaptation network (MS-MDA) to capture both domain-invariant and domain-specific features of emotional EEG signals. Further, Zhou \textit{et al.} \cite{zhou2023pr} proposed a novel prototypical representation-based pairwise learning framework (PR-PL) to address individual differences and noisy labeling in emotional EEG signals. Despite the progress made in dealing with individual differences, these methods struggle to mitigate the distributional differences across datasets, resulting in significant performance degradation in cross-corpus scenarios.}

\subsection{Cross-Corpus EEG Emotion Recognition}
\label{subsec:Cross-Corpus EEG-Based Emotion Recognition}{
To alleviate the large distributional differences across datasets, researchers have attempted to use domain adaptation techniques for cross-corpus EEG-based emotion recognition. For example, He \textit{et al.} \cite{he2022adversarial} proposed an adversarial discriminative temporal convolutional network (AD-TCN) that integrates the adversarial discriminative learning into a temporal convolutional network for enhancing distribution matching. Meanwhile, Li \textit{et al.} \cite{li2024distillation} proposed a novel distillation-based domain generalization network (DBDG) to learn the discriminative and generalizable emotional features. Moreover, Zhou \textit{et al.} \cite{zhou2025enhancing} introduced an EEG-based emotion style transfer network (E$^2$STN) that contains the content information of the source domain and the style information of the target domain, achieving superior performance in cross-corpus scenarios. However, these methods require prior access to all labeled source data and unlabeled target data for model training, which is not feasible in practical applications due to the difficulties in collecting and labeling EEG data. Therefore, recent studies have sought to leverage contrastive learning for cross-corpus EEG-based emotion recognition. A noteworthy attempt is JCFA \cite{liu2024joint}, which performs joint contrastive learning across three domains to align the time- and frequency-based embeddings of the same EEG sample in the latent time-frequency space, achieving SOTA performance in cross-corpus EEG-based emotion recognition tasks. However, such approaches following the traditional CL paradigm \cite{chen2020simple} fail to capture the emotional continuity inherent in EEG signals, resulting in relatively limited performance in cross-corpus scenarios. Therefore, this study aims to develop a sufficiently generalized model without relying on labeled data, which can be directly applied to different EEG emotion datasets without retraining from scratch, and achieve accurate and efficient cross-corpus generalization.}
}

\section{Problem Formulation}
\label{sec:Problem Formulation}{
Given an unlabeled pre-training EEG emotion dataset $X=\{x_i\}_{i=1}^{N}$ with $N$ samples, where each sample $x_i \in \mathbb{R}^{C \times F}$ contains $C$ channels and $F$-dimensional features, the goal is to learn a nonlinear embedding function $f_\theta$. This function is designed to map $x_i$ to its representation $h_i$ that best describes itself by leveraging the emotional continuity inherent in EEG signals. Ultimately, the pre-trained model is capable of producing generalizable EEG representations that can be effectively used across different datasets.
}

\begin{figure*}[h]
    \begin{center}
        \includegraphics[width=0.9\textwidth]{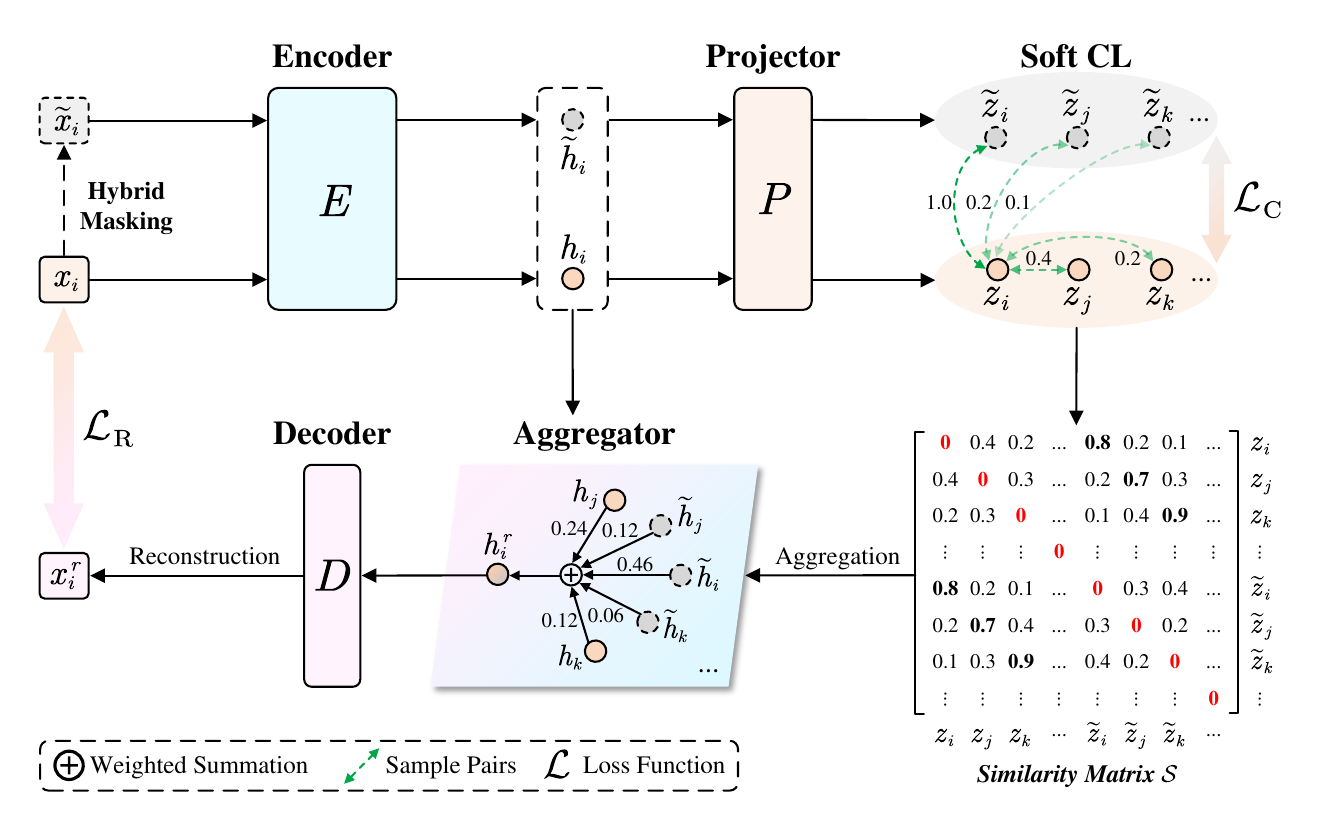}
    \end{center}
    \caption{The overall framework of SCMM. The pre-training process of SCMM involves three modules: (1) hybrid masking, (2) soft contrastive learning, and (3) aggregate reconstruction. Best viewed in color.}
    \label{fig:Overall Framework}
    \Description{The overall framework of SCMM.}
\end{figure*}

\section{Methodology}
\label{sec:Methodology}{
The overall framework of SCMM is illustrated in Fig. \ref{fig:Overall Framework}, which includes three main modules: hybrid masking, soft contrastive learning, and aggregate reconstruction. Below, we will detail the specific design of each module and the pre-training process of SCMM.

\subsection{Hybrid Masking}
\label{subsec:Hybrid Masking}{
The selection of masking strategies is crucial for CL and masked modeling. For an input EEG sample $x_i \in X$, existing methods use random masking \cite{zhang2022ganser} or channel masking \cite{li2022multi} to generate the masked sample $\widetilde{x}_i$. The random masking strategy masks samples along the feature dimension, ignoring the inter-channel relationships of multi-channel EEG signals. While a large masking ratio (e.g., 75\%) can mask entire portions of certain channels, it complicates the modeling process due to significant information loss. Conversely, the channel masking strategy masks features across all dimensions of the selected channels, losing the relationships between different dimensional features. Neither approach captures both channel and feature relationships simultaneously. Therefore, we develop a new hybrid masking strategy to generate diverse masked samples by considering both channel and feature relationships.

Specifically, we first generate a random masking matrix ${\rm Mask}_R \in \{0, 1\}$ with dimensions $C \times F$ and a channel masking matrix ${\rm Mask}_C \in \{0, 1\}$ with dimensions $C \times F$, both derived from binomial distributions with the same masking ratio $r \in (0, 1)$. Here, the element values in each row of ${\rm Mask}_C$ are either all 1s or all 0s. Next, we generate a probability matrix $U \in [0, 1]$ with dimensions $C \times 1$ for hybrid masking, which is drawn from a uniform distribution. The hybrid masking process is defined as:
\begin{equation}
\label{eq:Hybrid Masking}
    \widetilde{x}_{i, c} =
        \begin{cases}
            x_{i, c} \odot {\rm Mask}_{R, c} & \text{if } \mu < U_c \leq 1 \\
            x_{i, c} \odot {\rm Mask}_{C, c} & \text{if } 0 \leq U_c \leq \mu 
        \end{cases},
\end{equation}
where $\odot$ denotes element-wise multiplication. $x_{i,c}$ represents the DE features of the $c$-th channel of $x_i$, and $\widetilde{x}_{i,c}$ is the corresponding masked sample. $U_c$ represents the probability value of $U$ in the $c$-th row, and $\mu$ is a probability threshold that controls the weights of the two masking strategies. By integrating the hybrid masking strategy in SCMM, we enhance the diversity of masked samples, enabling the model to learn richer feature representations that account for both channel and feature relationships of EEG signals. Figure \ref{fig:Masking_Strategy} illustrates the differences between the three masking strategies. More details are presented in Appendix \ref{appendix:Masking Strategy}.

\begin{figure}[h]
    \begin{center}
        \includegraphics[width=0.475\textwidth]{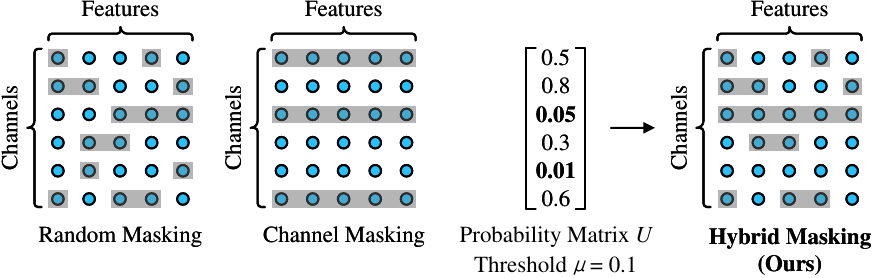}
    \end{center}
    \caption{Comparison of different masking strategies. The masking ratio and threshold are set to $r=0.5$ and $\mu=0.1$, respectively. Best viewed in color. Zoom in for a better view.}
    \label{fig:Masking_Strategy}
    \Description{Comparison of different masking strategies.}
\end{figure}

\subsection{Soft Contrastive Learning}
\label{subsec:Soft Contrastive Learning}{
Traditional hard CL treats the same sample and its augmented views as positive pairs, while treating all other samples as negatives \cite{chen2020simple}. During the computation of the contrastive loss, hard values (1 or 0) are assigned to sample pairs, as illustrated in Fig. \ref{fig:Illustration}(b) (Hard CL). However, we argue that this approach fails to account for the “short-term continuity” characteristic inherent in human emotions, leading to inaccurate modeling of inter-sample relationships and hindering the generalizability of the learned embeddings. 

To address this issue, we propose defining soft assignments for different sample pairs, as shown in Fig. \ref{fig:Illustration}(b) (SCMM). We first input $x_i$ and $\widetilde{x}_i$ into an encoder $E$ that maps samples to embeddings, denoted as $h_i = E(x_i)$ and $\widetilde{h}_i = E(\widetilde{x}_i)$. These embeddings are then projected into a latent space $\mathcal{Z}$ using a projector $P$, resulting in $z_i = P(h_i)$ and $\widetilde{z}_i = P(\widetilde{h}_i)$. Next, we perform soft contrastive learning in $\mathcal{Z}$ using $z_i$ and $\widetilde{z}_i$. Specifically, for a given pair of samples ($x_i$, $x_j$), we first calculate the normalized distance $D(x_i, x_j)$ between $x_i$ and $x_j$ in the original data space as:
\begin{equation}
\label{eq:Normalized Distance}
    D(x_i, x_j) = Norm(Dist(x_i, x_j)) \in [0, 1],
\end{equation}
where $Dist(\cdot, \cdot)$ is a metric function used to measure the distance between sample pairs, and $Norm(\cdot)$ denotes min-max normalization. Based on the normalized distance $D(x_i, x_j)$, we then define a soft assignment $w{(x_i, x_j)}$ for each pair of samples ($x_i$, $x_j$) using the sigmoid function $\sigma(x)=1/(1+{\rm exp}(-x))$:
\begin{equation}
\label{eq:Soft Assignments}
    w{(x_i, x_j)} = 2 \alpha \cdot \sigma(- D(x_i, x_j) / \tau_{\rm s}),
\end{equation}
where $\alpha \in [0, 1]$ is a boundary parameter that controls the upper bound of soft assignments. $\tau_{\rm s}$ is a sharpness parameter, where smaller values of $\tau_{\rm s}$ result in greater differences in $w(\cdot, \cdot)$ between sample pairs, and vice versa. Figure \ref{fig:Sharpness} illustrates the differences between soft assignments $w(\cdot, \cdot)$ with different sharpness $\tau_{\rm s}$.
\begin{figure}
    \begin{center}
        \includegraphics[width=0.475\textwidth]{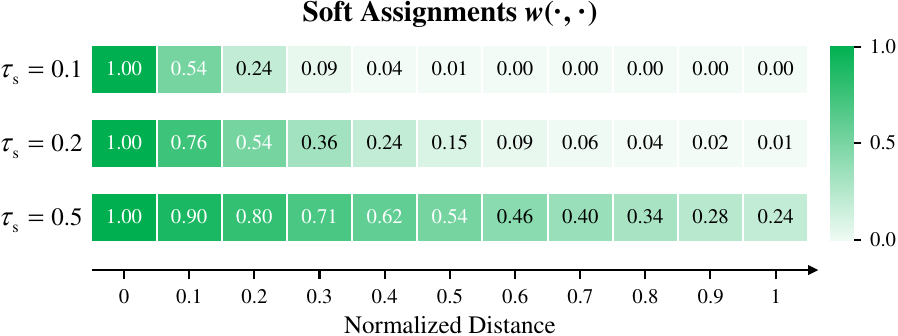}
    \end{center}
    \caption{Heatmaps of soft assignments $w(\cdot, \cdot)$ with different sharpness $\tau_{\rm s}$. Best viewed in color. Zoom in for a better view.}
    \label{fig:Sharpness}
    \Description{Soft assignments with different sharpness parameters.}
\end{figure}

Leveraging the soft assignments for all sample pairs, we propose a soft contrastive loss to refine the traditional hard contrastive loss. Specifically, for a pair of projected embeddings $(z_i, \widetilde{z}_i)$, we first calculate the softmax probability of the relative similarity among all similarities as:
\begin{equation}
\label{eq:Softmax Probability}
    p(z_i, \widetilde{z}_i) = \frac{{\rm exp}({\rm sim}(z_i, \widetilde{z}_i) / \tau_{\rm c})}{\sum_{z' \in \mathcal{Z} \backslash \{z_i\}} {\rm exp}({\rm sim}(z_i, z')/ \tau_{\rm c})},
\end{equation}
where ${\rm sim(\cdot, \cdot)}$ refers to the cosine similarity, and $\tau_{\rm c}$ is a temperature parameter used to adjust the scale. Based on $p(z_i, \widetilde{z}_i)$, the soft contrastive loss is then defined as:
\begin{equation}
\label{eq:Soft Contrastive Loss}
    \begin{split}
    \mathcal{L}_{{\rm C}, i} &= - {\rm log} \thinspace p(z_i, \widetilde{z}_i) \\
                       &- \sum_{\substack{{\bm x} \in \mathcal{X} \backslash \{x_i, \widetilde{x}_i\} \\ {\bm z} \in \mathcal{Z} \backslash \{z_i, \widetilde{z}_i\}}} w{(x_i, {\bm x})} \cdot {\rm log} \thinspace p(z_i, {\bm z}),
    \end{split}
\end{equation}
where $\mathcal{X} = X \cup \widetilde{X}$ represents the union of the data spaces of the original and masked samples. By assigning soft weights to different sample pairs, the model is encouraged to better capture the inherent correlations across different samples. During the training process, the final soft contrastive loss $\mathcal{L}_{\rm C}$ is computed by summing and averaging $\mathcal{L}_{{\rm C}, i}$ across all samples within a mini-batch. Notably, when $\forall{w(x_i, {\bm x})}=0$, the soft contrastive loss reduces to the traditional hard contrastive loss.
}

\subsection{Aggregate Reconstruction}
\label{subsec:Aggregate Reconstruction}{
To further capture the fine-grained relationships between different samples, we incorporate an aggregator for weighted aggregation and reconstruction. Current approaches for masked EEG modeling typically reconstruct the masked portion based on the unmasked portion of a single masked sample \cite{lan2024cemoae, pang2024multi}, following the learning paradigm of MAE \cite{he2022masked}. However, this single-sample reconstruction strategy overlooks the interactions between samples, leading to a complex and inadequate reconstruction process. 

To overcome this limitation, we introduce a similarity-aware aggregator that improves the traditional single-sample reconstruction process. Specifically, we first calculate the cosine similarity between each pair of projected embeddings $(z_i, z_j)$ within a mini-batch, resulting in a similarity matrix $\mathcal{S}$. Based on the pairwise similarities in $\mathcal{S}$, the aggregator then performs weighted aggregation of the embedding $h_i$, defined as:
\begin{equation}
\label{eq:Weighted Aggregation and Reconstruction}
    h_i^{r} = \sum_{z' \in \mathcal{Z} \backslash \{z_i\}} (\frac{{\rm exp} ({\rm sim}(z_i, z') / \tau_{\rm c})} {\sum_{z'' \in \mathcal{Z} \backslash \{z_i\}} {\rm exp} ({\rm sim}(z_i, z'') / \tau_{\rm c})} \cdot h'),
\end{equation}
where $h' \in \mathcal{H} \backslash \{h_i\}$ represents the encoded embedding corresponding to the projected embedding $z'$, and $\mathcal{H}$ denotes the embedding space of the encoder $E$. This approach allows for a more comprehensive reconstruction by aggregating complementary information and incorporating similar features from different samples during the reconstruction process. Finally, the reconstructed embedding $h_i^{r}$ is fed into a lightweight decoder $D$ to obtain the reconstructed sample $x_i^{r}$. Following the masked modeling paradigm, we use Mean Squared Error (MSE) as the reconstruction loss for model optimization, which is defined as:
\begin{equation}
\label{eq:Reconstruction Loss}
    \mathcal{L}_{{\rm R}, i} = \|x_i - x_i^{r}\|_2^2. 
\end{equation}

Similar to the soft contrastive loss $\mathcal{L}_{\rm C}$, the final reconstruction loss $\mathcal{L}_{\rm R}$ is computed by summing and averaging $\mathcal{L}_{{\rm R}, i}$ across all samples within a mini-batch. 
}

\begin{table*}[h]
    \caption{Experimental results on SEED and SEED-IV under two cross-corpus conditions: (1) same-class and (2) different-class. "$\dag$" and "$\ddag$" represent that the model uses DE features and raw EEG signals as inputs, respectively. "*" indicates that the results are reproduced by ourselves. A $\rightarrow$ B denotes that A is the pre-training dataset, while B is the dataset for model fine-tuning and testing. Best results are highlighted in bold, while the \underline{second-best results} are underlined.}
    \centering
    \begin{tabular}{c|c|c|c|c}
        \toprule
        \multirow{2.5}{*}{Methods} & \multicolumn{2}{c|}{\textbf{Same-Class}} & \multicolumn{2}{c}{\textbf{Different-Class}}\\
        \cmidrule(lr{0.01cm}){2-5} & SEED-IV$^3$ $\rightarrow$ SEED$^3$ & SEED$^3$ $\rightarrow$ SEED-IV$^3$ & SEED-IV$^4$ $\rightarrow$ SEED$^3$ & SEED$^3$ $\rightarrow$ SEED-IV$^4$ \\
        \midrule
           \multicolumn{5}{c}{\textbf{Transfer Learning}} \\
        \midrule
            DANN \cite{ganin2016domain}$^\dag$* & 51.91 / 09.27 & 45.90 / 03.60 & - & - \\
            BiDANN \cite{li2018novel}$^\dag$ & 49.24 / 10.49 & 60.46 / 11.17 & -  & - \\
            TANN \cite{li2021novel}$^\dag$ & 58.41 / 07.16 & 60.75 / 10.61 & -  & - \\
            PR-PL \cite{zhou2023pr}$^\dag$* & 61.01 / 10.55 & 58.74 / 10.71 & -  & - \\ 
            E$^2$STN \cite{zhou2025enhancing}$^\dag$ & 60.51 / 05.41 & 61.24 / 15.14 & - & - \\
        \midrule
           \multicolumn{5}{c}{\textbf{Self-Supervised Learning}} \\
        \midrule
            SimCLR \cite{chen2020simple,tang2020exploring}$^\ddag$* & 47.27 / 08.44 & 46.89 / 13.41 & 44.19 / 09.28 & 42.03 / 10.05 \\
            Mixup \cite{zhang2017mixup,wickstrom2022mixing}$^\ddag$* & 56.86 / 16.83 & 55.70 / 16.28 & 54.55 / 17.95 & 45.79 / 15.16 \\
            TS-TCC \cite{eldele2021time}$^\ddag$* & 55.38 / 11.65 & 49.43 / 09.44 & 52.30 / 12.49 & 44.57 / 06.08 \\
            MAE \cite{he2022masked}$^\dag$* & \underline{86.49 / 10.57} & \underline{83.87 / 08.53} & \underline{86.02 / 08.96} & \underline{76.74 / 09.18} \\
            JCFA \cite{liu2024joint}$^\ddag$ & 67.53 / 12.36 & 62.40 / 07.54 & 65.99 / 14.04 & 52.67 / 05.86 \\
        \midrule
            \rowcolor{dodgerblue!15}
            \textbf{SCMM (Ours)} & \textbf{91.61 / 07.56 {\color{red}(+05.12)}} & \textbf{87.24 / 08.35 {\color{red}(+03.37)}} & \textbf{91.26 / 07.91 {\color{red}(+05.24)}} & \textbf{80.89 / 08.69 {\color{red}(+04.15)}} \\
    \bottomrule
    \end{tabular}
\label{tab:Model performance on the SEED and SEED-IV datasets.}
\end{table*}

\subsection{The Pre-training Process of SCMM}
\label{subsec:Self-supervised Pre-training}{
During the pre-training process, SCMM is trained by jointly optimizing $\mathcal{L}_{\rm C}$ and $\mathcal{L}_{\rm R}$. The overall pre-training loss is defined as:
\begin{equation}
\label{eq:Pre-training Loss}
    \mathcal{L}_{\rm pret} = \lambda_{\rm C} \mathcal{L}_{\rm C} + \lambda_{\rm R} \mathcal{L}_{\rm R}, 
\end{equation}
where $\lambda_{\rm C}$ and $\lambda_{\rm R}$ are trade-off hyperparameters that are adaptively adjusted according to the homoscedastic uncertainty of each loss item \cite{kendall2018multi}. Algorithm \ref{alg:Pre-training Process of SCMM} details the pre-training process of SCMM.
\begin{algorithm}
    \caption{The pre-training process of SCMM.}
    \label{alg:Pre-training Process of SCMM}
    \begin{algorithmic}[1]
        \REQUIRE
    \renewcommand{\algorithmicrequire}{\textbf{}}
    \REQUIRE - Unlabeled pre-training EEG emotion dataset $X=\{x_i\}_{i=1}^{N}$. The number of pre-training \textit{epochs}.
    \ENSURE
    \STATE Randomly initialize the model parameters $\theta$; \\
        \FOR{$ epoch = 1 $ to \textit{epochs}}
            \item[] \textcolor{gray}{// All operations are performed within a mini-batch}
            \STATE Generate the masked sample $\widetilde{x}_i$ of each input EEG sample $x_i$ using hybrid masking in \textbf{Eq. (\ref{eq:Hybrid Masking})}; \\
            \STATE Generate $h_i$ and $\widetilde{h}_i$ by feeding $x_i$ and $\widetilde{x}_i$ into $E$;
            \STATE Generate $z_i$ and $\widetilde{z}_i$ by feeding $h_i$ and $\widetilde{h}_i$ into $P$;
            \STATE Compute the normalized distance $D(x_i, x_j)$ for each pair of samples $(x_i, x_j)$ using \textbf{Eq. (\ref{eq:Normalized Distance})};
            \STATE Generate the soft assignment $w(x_i, x_j)$ for each pair of samples $(x_i, x_j)$ using \textbf{Eq. (\ref{eq:Soft Assignments})};
            \STATE Compute the soft contrastive loss $\mathcal{L}_{\rm C}$ using \textbf{Eq. (\ref{eq:Soft Contrastive Loss})}; \\
            \STATE Compute the pairwise cosine similarity for each pair of projected embeddings $(z_i, z_j)$; 
            \STATE Generate the reconstructed embedding $h_i^{r}$ of each $h_i$ through weighted aggregation in \textbf{Eq. (\ref{eq:Weighted Aggregation and Reconstruction})};
            \STATE Reconstruct $x_i^{r}$ by feeding $h_i^{r}$ into $D$;
            \STATE Compute the reconstruction loss $\mathcal{L}_{\rm R}$ using \textbf{Eq. (\ref{eq:Reconstruction Loss})}; \\
            \STATE Compute the pre-training loss $\mathcal{L}_{\rm pret}$ using \textbf{Eq. (\ref{eq:Pre-training Loss})};
            \STATE Update the model parameters $\theta$;
        \ENDFOR \\
    \RETURN The pre-trained SCMM model $f_\theta$. \\
    \end{algorithmic}
\end{algorithm}}
}

\section{Experiments}
\label{sec:Experiments}{
\subsection{Datasets}
\label{subsec:Datasets}{
We conduct extensive experiments on three public datasets, SEED \cite{zheng2015investigating}, SEED-IV \cite{zheng2018emotionmeter}, and DEAP \cite{koelstra2011deap}, to evaluate the model performance of SCMM in cross-corpus EEG-based emotion recognition tasks. These datasets are diverse in terms of EEG equipment, emotional stimuli, data specifications, labeling approaches, and subjects, making them well-suited for assessing the model's efficacy in cross-corpus scenarios. In the experiments, we use 1-s (SEED and DEAP) and 4-s (SEED-IV) DE features as inputs, respectively. Detailed descriptions of the datasets and pre-processing procedures are provided in Appendix \ref{appendix:Datasets}.
}

\subsection{Implementation Details}
\label{subsec:Implementation Details}{
In the pre-training stage, we set $r$ to 0.5 and $\mu$ to 0.1 for hybrid masking. We use the negative of cosine similarity as $D(\cdot, \cdot)$, and we set $\alpha$ to 0.5, $\tau_{\rm s}$ to 0.05, and $\tau_{\rm c}$ to 0.5 for soft CL. We use Adam as optimizer with an initial learning rate of 5e-4 and a weight decay of 3e-4. The pre-training process is conducted over 200 epochs with a batch size of 256. We save the model parameters $\theta$ from the final epoch as the pre-trained SCMM. In the fine-tuning stage, we input the encoded embedding $h_i$ into an emotion classifier consisting of a 2-layer fully connected network for final emotion recognition. For efficient deployment and testing of the model, the pre-trained SCMM is optimized solely using cross-entropy loss during fine-tuning. The fine-tuning process is conducted over 50 epochs with a batch size of 128. All experiments are conducted using Python 3.9 with PyTorch 1.13 on an NVIDIA GeForce RTX 3090 GPU. We release the source code of SCMM at https://github.com/Kyler-RL/SCMM. Further implementation details can be found in Appendix \ref{appendix:Implementation Details}.
}

\subsection{Baseline Models and Experimental Settings}
\label{subsec:Baseline Models and Experimental Settings}{
We compare the proposed SCMM against ten competitive baselines, including five transfer learning methods: DANN \cite{ganin2016domain}, BiDANN \cite{li2018novel}, TANN \cite{li2021novel}, PR-PL \cite{zhou2023pr}, and E$^2$STN \cite{zhou2025enhancing}, as well as five self-supervised learning models: SimCLR \cite{chen2020simple, tang2020exploring}, Mixup \cite{zhang2017mixup, wickstrom2022mixing}, TS-TCC \cite{eldele2021time}, MAE \cite{he2022masked}, and JCFA \cite{liu2024joint}. Notably, E$^2$STN and JCFA are two SOTA methods designed for cross-corpus EEG-based emotion recognition. In the experiments, we adopt a cross-corpus subject-independent protocol consistent with JCFA and use a leave-trials-out cross-validation strategy for fine-tuning and testing. We calculate the average accuracy and standard deviation (ACC / STD $\%$) across all subjects in the test set to evaluate the model performance of SCMM. More details about baseline models and experimental settings are provided in Appendix \ref{appendix:Experimental Settings}.
}

\begin{table*}[h]
    \caption{Experimental results on SEED and DEAP under the different-class cross-corpus condition.}
    \centering
    \begin{tabular}{c|c|c|c}
        \toprule
        Methods & DEAP $\rightarrow$ SEED$^3$ & SEED$^3$ $\rightarrow$ DEAP (Valence) & SEED$^3$ $\rightarrow$ DEAP (Arousal) \\ 
        \midrule
            SimCLR \cite{chen2020simple,tang2020exploring}$^\ddag$* & 53.12 / 13.12 & 53.75 / 03.61 & 51.79 / 04.54 \\
            Mixup \cite{zhang2017mixup,wickstrom2022mixing}$^\ddag$* & 48.75 / 14.37 & 60.62 / 08.68 & 60.11 / 07.69 \\
            TS-TCC \cite{eldele2021time}$^\ddag$* & 49.37 / 12.50 & 56.25 / 03.46 & 54.13 / 04.45 \\
            MAE \cite{he2022masked}$^\dag$* & \underline{83.69 / 10.10} & \underline{72.19 / 07.24} & \underline{70.50 / 06.30} \\
            JCFA \cite{liu2024joint}$^\ddag$ & 64.69 / 12.28 & 61.59 / 06.26 & 61.06 / 07.37 \\
        \midrule
            \rowcolor{dodgerblue!15}
            \textbf{SCMM (Ours)} & \textbf{91.70 / 08.07 {\color{red}(+08.01)}} & \textbf{73.96 / 06.75 {\color{red}(+01.77)}} & \textbf{72.66 / 05.67 {\color{red}(+02.16)}} \\
    \bottomrule
    \end{tabular}
\label{tab:Model performance on SEED and DEAP.}
\end{table*}

\begin{table*}[h]
    \caption{Ablation study on SEED and SEED-IV under same-class and different-class cross-corpus conditions.}
    \centering
    \begin{tabular}{c|c|c|c|c}
        \toprule
        \multirow{2.5}{*}{Methods} & \multicolumn{2}{c|}{\textbf{Same-Class}} & \multicolumn{2}{c}{\textbf{Different-Class}} \\
        \cmidrule(lr{0.01cm}){2-5}
        & SEED-IV$^3$ $\rightarrow$ SEED$^3$ & SEED$^3$ $\rightarrow$ SEED-IV$^3$ & SEED-IV$^4$ $\rightarrow$ SEED$^3$ & SEED$^3$ $\rightarrow$ SEED-IV$^4$ \\
        \midrule
        w/o $\mathcal{L}_{\rm C}$ & 89.68 / 09.32 & 84.24 / 11.90 & 89.45 / 09.10 & 77.24 / 09.14 \\
        w/o $\mathcal{L}_{\rm R}$ & 90.73 / 08.48 & 85.07 / 11.05 & 90.96 / 08.36 & 78.32 / 07.19 \\
        \rowcolor{dodgerblue!15}
        \textbf{SCMM} & \textbf{91.61 / 07.56} & \textbf{87.24 / 08.35} & \textbf{91.26 / 07.91} & \textbf{80.89 / 08.69} \\
    \bottomrule
    \end{tabular}
\label{tab:Ablation study on the SEED and SEED-IV datasets.}
\end{table*}

\subsection{Results Analysis and Comparison}
\label{subsec:Results Analysis and Comparison}{
To fully validate the model performance of SCMM, we conduct extensive experiments under two cross-corpus conditions: (1) \textbf{same-class} and (2) \textbf{different-class}. Appendix \ref{appendix:Within-Dataset Validation} presents the within-dataset validation experimental results of SCMM on SEED and SEED-IV. Full results are provided in Appendix \ref{appendix:Full Results}.

(1) \textbf{Same-Class.} We first conduct two experiments on the SEED and SEED-IV 3-category datasets: pre-training on SEED-IV and fine-tuning on SEED (SEED-IV$^3$ $\rightarrow$ SEED$^3$), and pre-training on SEED and fine-tuning on SEED-IV (SEED$^3$ $\rightarrow$ SEED-IV$^3$). In both experiments, all samples corresponding to fear emotions in the SEED-IV dataset are excluded. The left two columns in Table \ref{tab:Model performance on the SEED and SEED-IV datasets.} present the comparison results, indicating that SCMM achieves SOTA performance in both experiments. Specifically, our model achieves classification accuracies of 91.61\% and 87.24\% with standard deviations of 7.56\% and 8.35\% in the SEED-IV$^3$ $\rightarrow$ SEED$^3$ and SEED$^3$ $\rightarrow$ SEED-IV$^3$ experiments, outperforming the second-best method MAE by accuracies of 5.12\% and 3.37\%, respectively. In addition, the proposed SCMM is significantly better than transfer learning methods, highlighting its superiority.

(2) \textbf{Different-Class.} We then conduct two experiments on the SEED and SEED-IV 4-category datasets, denoted as SEED-IV$^4$ $\rightarrow$ SEED$^3$ and SEED$^3$ $\rightarrow$ SEED-IV$^4$. These experiments aim to evaluate the model performance when the pre-training and fine-tuning datasets contain different emotion categories. In the experiments, we exclude transfer learning methods since they are not suitable for scenarios where the training and testing datasets contain different emotion categories. Experimental results in the right two columns of Table \ref{tab:Model performance on the SEED and SEED-IV datasets.} demonstrate that SCMM achieves the best performance in both experiments. Specifically, our model achieves classification accuracies of 91.26\% and 80.89\% with standard deviations of 7.91\% and 8.69\% in the SEED-IV$^4$ $\rightarrow$ SEED$^3$ and SEED$^3$ $\rightarrow$ SEED-IV$^4$ experiments, surpassing the second-best method MAE by 5.24\% and 4.15\% in accuracies, respectively. Additionally, the traditional CL-based models SimCLR, Mixup, TS-TCC, and JCFA exhibit relatively poor performance across all experiments due to their use of raw EEG signals as inputs.

To further validate the generalization capability of SCMM, we conduct additional experiments on the SEED and DEAP datasets, denoted as DEAP $\rightarrow$ SEED$^3$, SEED$^3$ $\rightarrow$ DEAP (Valence), and SEED$^3$ $\rightarrow$ DEAP (Arousal). Note that the EEG acquisition equipment, emotional stimuli, data specifications, labeling approaches, and subjects are completely different between the two datasets. Table \ref{tab:Model performance on SEED and DEAP.} presents the experimental results of SCMM compared to existing methods. Specifically, for the DEAP $\rightarrow$ SEED$^3$ experiment, SCMM achieves an accuracy of 91.70\% with a standard deviation of 8.07\%, outperforming the second-best method MAE by an accuracy of 8.01\%. For the SEED$^3$ $\rightarrow$ DEAP (Valence) and SEED$^3$ $\rightarrow$ DEAP (Arousal) experiments, SCMM achieves classification accuracies of 73.96\% and 72.66\% with standard deviations of 6.75\% and 5.67\%, surpassing the second-best method MAE by 1.77\% and 2.16\% in accuracy. The results show that our model maintains excellent performance even when the pre-training and fine-tuning datasets are completely different, highlighting its superior generalization capability. Further, the comparative analysis of Table \ref{tab:Model performance on the SEED and SEED-IV datasets.} and Table \ref{tab:Model performance on SEED and DEAP.} reveals that the model performance of SCMM on the same fine-tuning dataset remains stable when pre-training on different datasets. This suggests that our model effectively captures generalizable emotional EEG representations that are robust to dataset variations.

In summary, extensive experimental results on SEED, SEED-IV, and DEAP confirm that our model exhibits superior performance and stability in cross-corpus EEG-based emotion recognition tasks under both same-class and different-class conditions.}
}

\begin{table*}[h]
    \caption{Comparison of hard CL and soft CL on SEED and SEED-IV under same-class and different-class cross-corpus conditions.}
    \centering
    \begin{tabular}{c|c|c|c|c}
        \toprule
        \multirow{2.5}{*}{Methods} & \multicolumn{4}{c}{ACC / STD (\%)} \\ 
        \cmidrule(lr{0.01cm}){2-5} & SEED-IV$^3$ $\rightarrow$ SEED$^3$ & SEED$^3$ $\rightarrow$ SEED-IV$^3$ & SEED-IV$^4$ $\rightarrow$ SEED$^3$ & SEED$^3$ $\rightarrow$ SEED-IV$^4$ \\
        \midrule
            Hard CL & 90.30 / 07.94 & 85.95 / 08.74 & 90.91 / 08.61 & 79.82 / 07.00 \\
            \rowcolor{dodgerblue!15}
            \textbf{Soft CL} & \textbf{91.61 / 07.56} & \textbf{87.24 / 08.35} & \textbf{91.25 / 07.91} & \textbf{80.89 / 08.69} \\
        \midrule & DEAP $\rightarrow$ SEED$^3$ & SEED$^3$ $\rightarrow$ DEAP (Valence) & SEED$^3$ $\rightarrow$ DEAP (Arousal) & - \\
        \midrule
            Hard CL & 90.87 / 08.58 & 73.10 / 07.34 & 71.98 / 06.02 & - \\
            \rowcolor{dodgerblue!15}
            \textbf{Soft CL} & \textbf{91.70 / 08.01} & \textbf{73.96 / 06.75} & \textbf{72.66 / 05.67} & - \\
        \bottomrule
    \end{tabular}
\label{tab:Comparison of hard CL and soft CL.}
\end{table*}

\begin{table*}[h]
    \caption{Comparison of soft CL in the embedding space (ES) and original data space (OS) on SEED and SEED-IV under same-class and different-class cross-corpus conditions.}
    \centering
    \begin{tabular}{c|c|c|c|c}
        \toprule
        \multirow{2.5}{*}{Methods} & \multicolumn{4}{c}{ACC / STD (\%)} \\ 
        \cmidrule(lr{0.01cm}){2-5} & SEED-IV$^3$ $\rightarrow$ SEED$^3$ & SEED$^3$ $\rightarrow$ SEED-IV$^3$ & SEED-IV$^4$ $\rightarrow$ SEED$^3$ & SEED$^3$ $\rightarrow$ SEED-IV$^4$ \\
        \midrule
            ES & 89.99 / 10.25 & 85.75 / 14.00 & 90.31 / 08.59 & 79.04 / 06.95 \\
            \rowcolor{dodgerblue!15}
            \textbf{OS} & \textbf{91.61 / 07.56} & \textbf{87.24 / 08.35} & \textbf{91.25 / 07.91} & \textbf{80.89 / 08.69} \\
        \midrule & DEAP $\rightarrow$ SEED$^3$ & SEED$^3$ $\rightarrow$ DEAP (Valence) & SEED$^3$ $\rightarrow$ DEAP (Arousal) & - \\
        \midrule
             ES & 90.64 / 07.97 & 72.75 / 07.06 & 71.58 / 05.72 & - \\
             \rowcolor{dodgerblue!15}
             \textbf{OS} & \textbf{91.70 / 08.01} & \textbf{73.96 / 06.75} & \textbf{72.66 / 05.67} & - \\
        \bottomrule
    \end{tabular}
\label{tab:Comparison of soft CL in the embedding space (ES) and original data space (OS).}
\end{table*}

\section{Discussions}
\label{sec:Discussions}{
\subsection{Ablation Study}
\label{subsec:Ablation Study}{
To assess the validity of each module in SCMM, we conduct a comprehensive ablation study on the SEED and SEED-IV datasets. Table \ref{tab:Ablation study on the SEED and SEED-IV datasets.} presents the results of ablation experiments. Specifically, we design three different models below. (1) \textbf{w/o $\mathcal{L}_{\rm C}$}: the first configuration removes the soft contrastive loss and trains the model using only the reconstruction loss. The results show that the model performs the worst without the contrastive learning constraint. However, it still outperforms the MAE model based on the single-sample reconstruction paradigm. This suggests that our aggregate reconstruction strategy effectively improves the model performance by capturing fine-grained inter-sample relationships. (2) \textbf{w/o $\mathcal{L}_{\rm R}$}: the second configuration removes the reconstruction loss and trains the model using only the soft contrastive loss. Experimental results show that this configuration outperforms the masked modeling approach, indicating that contrastive learning is more effective for EEG-based emotion recognition by capturing high-level semantic features of EEG signals. (3) \textbf{SCMM}: the last configuration trains the model with the soft contrastive loss and reconstruction loss, representing the full SCMM model. The results demonstrate that this configuration achieves the best performance in all experiments, indicating that SCMM significantly enhances the model performance and stability by combining soft contrastive learning and aggregate reconstruction. This improvement is evident under different cross-corpus conditions, demonstrating the feasibility of extending SCMM to real-life aBCI applications.
}

\begin{figure}
    \begin{center}
        \includegraphics[width=0.475\textwidth]{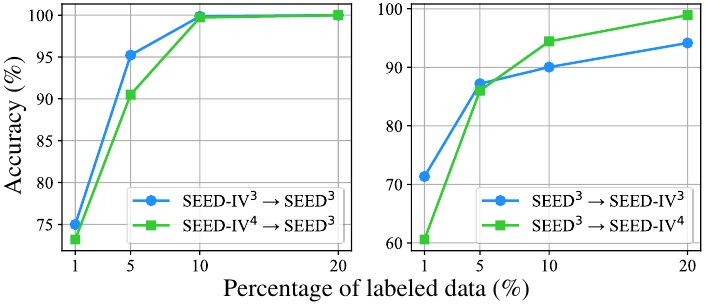}
    \end{center}
    \caption{Model performance with limited labeled data for fine-tuning on SEED and SEED-IV under same-class and different-class cross-corpus conditions. Best viewed in color. Zoom in for a better view.}
    \label{fig:Percentage of Labeled Data}
    \Description{Model performance with limited labeled data for fine-tuning on SEED and SEED-IV.}
\end{figure}

\subsection{Model Performance with Limited Data}
\label{subsec:Model Performance with Limited Data}{
We investigate the model performance of SCMM on the SEED and SEED-IV datasets when fine-tuning with limited labeled data. Specifically, we randomly select 1\%, 5\%, 10\%, and 20\% of labeled samples from the fine-tuning dataset for model fine-tuning, while the remaining samples are used for testing. Figure \ref{fig:Percentage of Labeled Data} shows the classification accuracy curves. For the SEED-IV$^3$ $\rightarrow$ SEED$^3$ and SEED-IV$^4$ $\rightarrow$ SEED$^3$ experiments, SCMM achieves classification accuracies exceeding 70\% with only 1\% of labeled data. The accuracies significantly improve as the proportion of labeled samples increases, reaching close to 100\% with 10\% of labeled data. Meanwhile, our model achieves classification accuracies over 60\% and 70\% with only 1\% of labeled data in the SEED$^3$ $\rightarrow$ SEED-IV$^3$ and SEED$^3$ $\rightarrow$ SEED-IV$^4$ experiments. The accuracies exceed 90\% when fine-tuning with 10\% of labeled samples in both experiments. In summary, the results indicate that SCMM maintains superior performance even with limited labeled data for fine-tuning, showing its outstanding robustness and potential in few-shot scenarios.
}

\begin{figure*}[h]
    \begin{center}
        \includegraphics[width=0.9\textwidth]{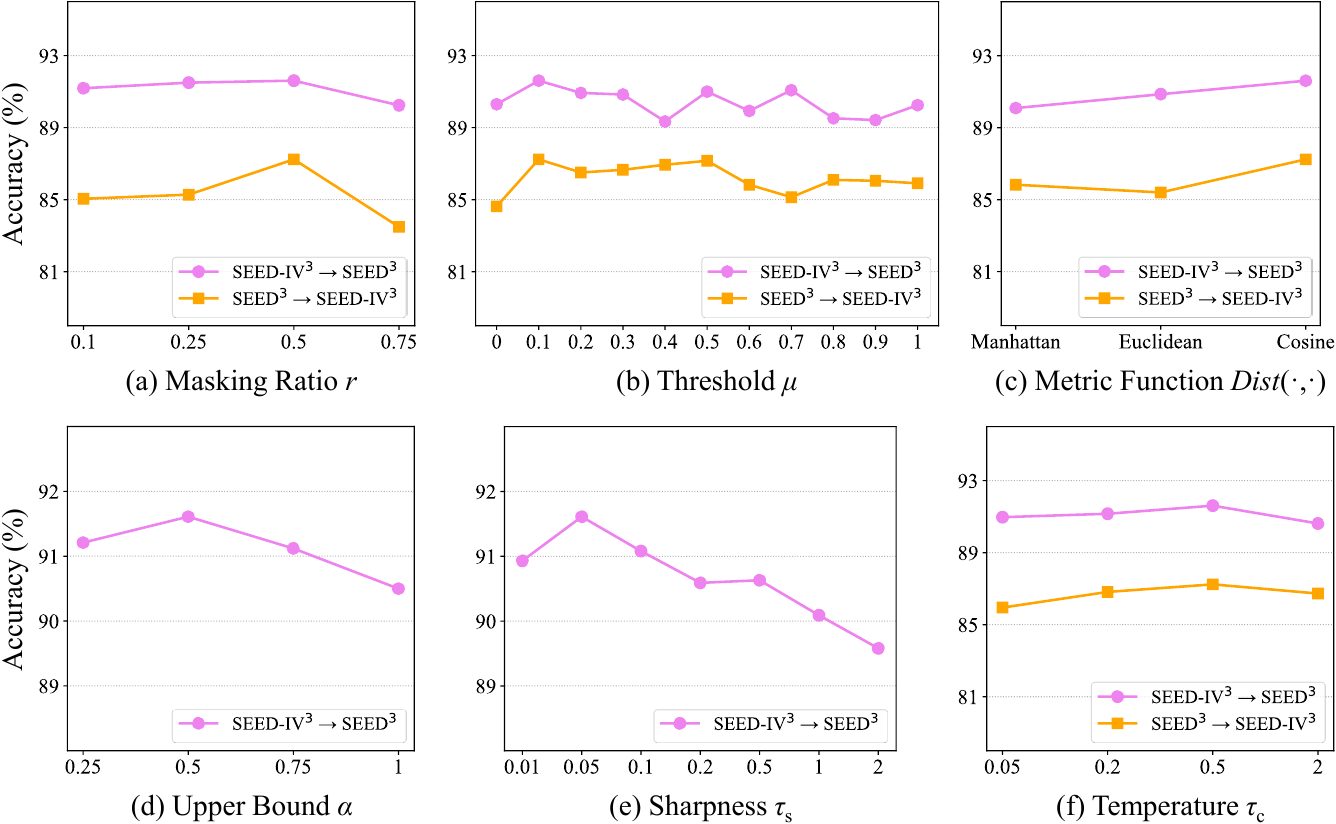}
    \end{center}
    \caption{Hyperparameter sensitivity analysis. (a) - (f) represent the masking ratio $r$, threshold $\mu$, metric function $Dist(\cdot, \cdot)$, upper bound $\alpha$, sharpness $\tau_{\rm s}$, and temperature $\tau_{\rm c}$, respectively. Best viewed in color.}
    \label{fig:Sensitivity_Analysis}
    \Description{Sensitivity analysis of hyperparameters.}
\end{figure*}

\subsection{Delve into Contrastive Learning}
\label{subsec:Delve into Soft Contrastive Learning}{
\subsubsection{Hard CL vs. Soft CL}{
We compare the model performance of SCMM using traditional hard CL and soft CL on SEED, SEED-IV, and DEAP under same-class and different-class cross-corpus conditions to verify the effectiveness of the proposed soft weighting mechanism. Table \ref{tab:Comparison of hard CL and soft CL.} presents the comparison results, showing that the SCMM model using the soft weighting mechanism consistently achieves the highest classification accuracies and lowest standard deviations in all experiments. This improvement indicates that assigning soft weights (rather than hard values) to different sample pairs when computing the contrastive loss effectively captures the inter-sample similarity relationships, thereby improving the discriminative ability of the model.
}

\subsubsection{Soft CL in the Embedding Space vs. Original Data Space}{While soft CL has been explored across various domains, most methods focus on computing soft assignments for contrastive loss in the embedding space \cite{dwibedi2021little, yeche2021neighborhood}. However, we argue that utilizing similarities in the original data space can provide better self-supervision and is particularly suitable for emotional EEG data. To validate this, we conduct experiments on SEED, SEED-IV, and DEAP under same-class and different-class cross-corpus conditions to compare soft CL in different spaces. Specifically, we modify the metric function $Dist(\cdot, \cdot)$ to use similarities between projected embeddings, shifting the computation of soft assignments from the original data space to the embedding space. Table \ref{tab:Comparison of soft CL in the embedding space (ES) and original data space (OS).} presents the experimental results, demonstrating that soft CL in the original data space consistently outperforms the embedding space in all experiments. Furthermore, this approach allows offline pre-computation of cosine similarities of different sample pairs in the original data space, thus reducing computational costs and improving training efficiency.}
}

\subsection{Hyperparameter Sensitivity Analysis}
\label{subsec:Hyperparameter Sensitivity Analysis}{
To analyze the hyperparameter sensitivity of SCMM, we conduct experiments on SEED and SEED-IV under the same-class condition, as shown in Fig. \ref{fig:Sensitivity_Analysis}. Specifically, the examined hyperparameters are divided into two groups: \textbf{(1) Hybrid Masking:} masking ratio $r$ and threshold $\mu$, and \textbf{(2) Soft Contrastive Learning:} metric function $Dist(\cdot, \cdot)$, upper bound $\alpha$, sharpness $\tau_{\rm s}$, and temperature $\tau_{\rm c}$.

(1) \textbf{Hybrid Masking.} Figures \ref{fig:Sensitivity_Analysis}(a) - (b) show the classification curves of SCMM with different masking ratios $r$ and thresholds $\mu$. Experimental results show that the proposed SCMM achieves the best performance when the masking ratio and threshold are set to $r=0.5$ and $\mu=0.1$ (i.e., the ratio of random masking and channel masking is 9:1), respectively. In addition, we find that our model performs better when using hybrid masking compared to random masking or channel masking in most settings, demonstrating the effectiveness of the proposed hybrid masking strategy in enhancing the model performance.

(2) \textbf{Soft Contrastive Learning.} Figures \ref{fig:Sensitivity_Analysis}(c) - (f) depict the classification curves of SCMM using different metric function $Dist(\cdot,\cdot)$, upper bound $\alpha$, sharpness $\tau_{\rm s}$, and temperature $\tau_{\rm c}$, respectively. Specifically, our model performs best on the SEED and SEED-IV 3-category datasets when using cosine similarity as the metric function $Dist(\cdot,\cdot)$, as shown in Fig. \ref{fig:Sensitivity_Analysis}(c). Additionally, Figures \ref{fig:Sensitivity_Analysis}(d) - (f) show that SCMM achieves the best performance when the upper bound, sharpness and temperature are set to $\mu=0.5$, $\tau_{\rm s}=0.05$ and $\tau_{\rm c}=0.5$, respectively. In summary, the results demonstrate that SCMM maintains excellent performance under different hyperparameter settings, indicating that its superior generalization ability is not significantly affected by hyperparameter changes.
}

\subsection{Visualization}
\label{subsec:Visualization}{
\subsubsection{Sample-Wise Relationships}{
To evaluate whether sample-wise relationships are preserved in the encoder, we randomly select 100 test samples from the SEED dataset and visualize the pairwise cosine similarity between sample pairs. Additionally, we select all test samples of one subject from the SEED dataset and visualize the learned embeddings of SCMM using t-SNE \cite{van2008visualizing}. Figure \ref{fig:Visualization}(a) presents heat maps of pairwise similarity matrices, where darker colors indicate higher similarity between samples. Traditional hard CL identifies only coarse-grained relationships across samples from different emotion categories, especially for the most challenging-to-recognize negative and neutral emotions. In contrast, SCMM effectively captures the fine-grained relationships between samples of different categories. Moreover, the results of t-SNE visualization in Fig. \ref{fig:Visualization}(b) indicate that our model better clusters samples within the same category and increases the inter-class distance compared to hard CL, thus enhancing the classification performance.
\begin{figure}[h]
    \begin{center}
        \includegraphics[width=0.475\textwidth]{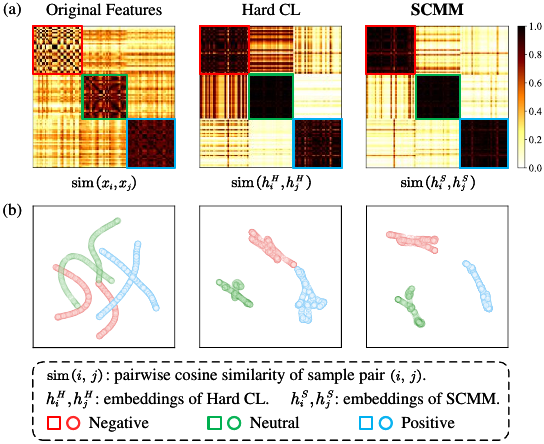}
    \end{center}
    \caption{(a) Heat maps of pairwise similarity matrices. (b) t-SNE visualization of the learned embeddings. Best viewed in color. Zoom in for a better view.}
    \label{fig:Visualization}
    \Description{Visualization.}
\end{figure}

\subsubsection{Intra- and Inter-Class Similarities}{
\begin{figure*}[h]
    \begin{center}
        \includegraphics[width=0.85\textwidth]{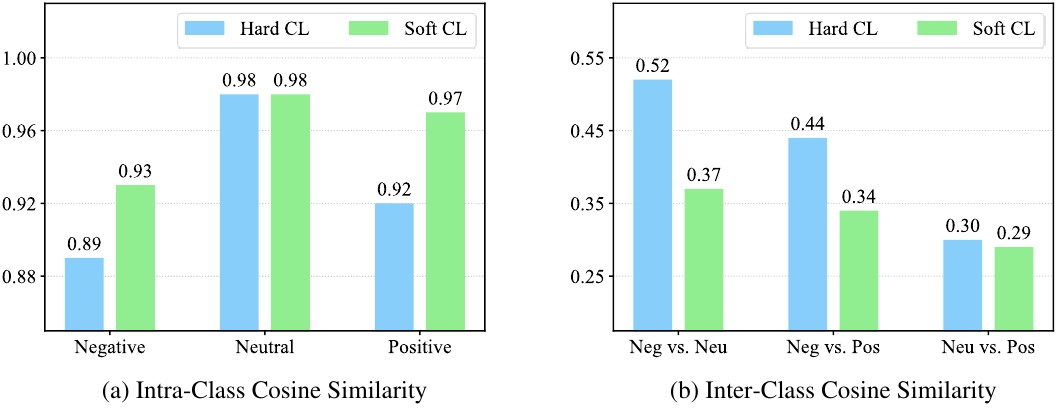}
    \end{center}
    \caption{Intra- and inter-class cosine similarities of embeddings learned by hard CL and soft CL. (a) and (b) represent the average intra-class and inter-class cosine similarities, respectively. Best viewed in color.}
    \label{fig:Intra_Inter_Similarity}
    \Description{Intra- and inter-class cosine similarities.}
\end{figure*}

\begin{figure*}[h]
    \begin{center}
        \includegraphics[width=0.85\textwidth]{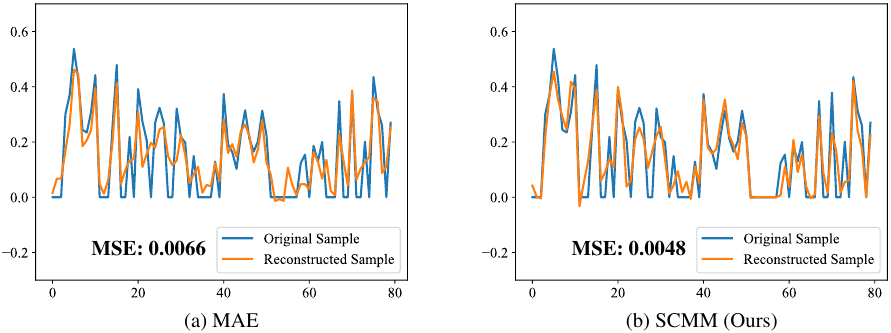}
    \end{center}
    \caption{Comparison of reconstruction quality. We visualize the reconstruction results to compare the single-sample reconstruction paradigm (MAE) with the aggregate reconstruction paradigm (SCMM). Best viewed in color.}
    \label{fig:Reconstruction}
    \Description{Comparison of reconstruction quality.}
\end{figure*}

To assess the quality of the embeddings learned by SCMM, we randomly select one subject from the SEED dataset and calculate both the average intra- and inter-class cosine similarities between the learned embeddings of all test samples, as shown in Fig. \ref{fig:Intra_Inter_Similarity}. It is evident that the proposed SCMM produces embeddings with higher intra-class similarity compared to traditional hard CL. In addition, the average inter-class similarity of the embeddings learned by SCMM is significantly lower than that of hard CL. In summary, visualization results confirm that the soft contrastive learning strategy designed in SCMM effectively clusters samples within the same category while distinctly separating samples from different categories, thus enhancing the model's discriminative capabilities.}

\subsubsection{Reconstruction Quality}{
To verify the effectiveness of the similarity-aware aggregator designed in SCMM, we compare the reconstruction quality of the single-sample reconstruction paradigm (MAE) with the aggregate reconstruction paradigm (SCMM) on the DEAP dataset. For clarity, we flatten both the original input sample and the reconstructed sample into one-dimensional vectors with dimensions $C \times F$ (channels $\times$ features). The results depicted in Fig. \ref{fig:Reconstruction} illustrate that our model achieves lower reconstruction loss (MSE) and better sample reconstruction, which is crucial to improving the model performance.}}
}

\section{Conclusions}
\label{sec:Conclusions}{
This paper proposes a novel SCMM framework to tackle the critical challenge of cross-corpus generalization in the field of EEG-based emotion recognition. Unlike traditional CL-based models, SCMM integrates soft contrastive learning with a hybrid masking strategy to effectively capture the "short-term continuity" characteristics inherent in human emotions and produce stable and generalizable EEG representations. Additionally, a similarity-aware aggregator is introduced to fuse complementary information from semantically related samples, thereby enhancing the fine-grained feature representation capability of the model. Extensive experimental results on three well-recognized datasets show that SCMM consistently achieves SOTA performance in cross-corpus EEG-based emotion recognition tasks under both same-class and different-class conditions. Comprehensive ablation study and hyperparameter sensitivity analysis confirm the superiority and robustness of SCMM. Visualization results indicate that our model effectively reduces the distance between similar samples within the same category and captures more fine-grained relationships across samples. These findings suggest that SCMM enhances the feasibility of extending the proposed method to real-life aBCI applications.}
}

\section{Acknowledgements}
\label{sec:Acknowledgements}{
This work was funded by the National Natural Science Foundation of China (62276169), the STI 2030-Major Projects (2021ZD0200500), the Medical-Engineering Interdisciplinary Research Foundation of Shenzhen University (2023YG004), the Shenzhen-Hong Kong Institute of Brain Science-Shenzhen Fundamental Research Institutions (2023SHIBS0003), the Open Research Fund of the State Key Laboratory of Brain-Machine Intelligence, Zhejiang University (Grant No. BMI2400008), and the Shenzhen Science and Technology Program (No. JCYJ20241202124222027 and JCYJ20241202124209011).
}

\bibliographystyle{ACM-Reference-Format}
\bibliography{acmart}

\appendix

\section{Datasets}
\label{appendix:Datasets}{
\subsection{Dataset Description}{
We conduct extensive experiments on three well-known datasets, SEED \cite{zheng2015investigating}, SEED-IV \cite{zheng2018emotionmeter}, and DEAP \cite{koelstra2011deap}, to evaluate the model performance of SCMM in cross-corpus EEG-based emotion recognition tasks. These datasets encompass different EEG equipment, emotional stimuli, data specifications, labeling approaches, and subjects, making them exceptionally suitable for evaluating the model's efficacy in cross-corpus scenarios. Table \ref{tab:Datasets} provides a detailed description of the three datasets.

\begin{table*}[h]
    \caption{Detailed descriptions of the experimental datasets.}
    \centering
    \begin{tabular}{c|c|c|c|c|c}
        \toprule
            Datasets & Subjects & Sessions $\times$ Trials & Channels & Sampling Rate & Classes \\
            \midrule
            SEED \cite{zheng2015investigating} & 15 & 3 $\times$ 15 & 62 & 1 kHz & 3 (Negative, Neutral, Positive)\\
            SEED-IV \cite{zheng2018emotionmeter} & 15 & 3 $\times$ 24 & 62 & 1 kHz & 4 (Sad, Neutral, Fear, Happy)\\ 
            DEAP \cite{koelstra2011deap} & 32 & 1 $\times$ 40 & 32 & 512 Hz & Valence: 1 - 9, Arousal: 1 - 9\\ 
        \bottomrule
    \end{tabular}
\label{tab:Datasets}
\end{table*}

(1) \textbf{SEED} \cite{zheng2015investigating} was developed by the Center for Brain-like Computing and Machine Intelligence (BCMI) of Shanghai Jiao Tong University. The dataset used a 62-channel ESI NeuroScan System based on the international 10-20 system to record EEG signals from 15 subjects (7 males and 8 females) under different video stimuli at a sampling rate of 1 kHz. Each subject participated in 3 sessions. In each session, each subject was required to watch 15 movie clips consisting of 3 different emotional states: negative, neutral, and positive. Each emotional state contains a total of 5 movie clips, corresponding to 5 trials.

(2) \textbf{SEED-IV} \cite{zheng2018emotionmeter} used the same EEG equipment as the SEED dataset, but with different video stimuli, emotion categories, and subjects. The dataset recorded EEG signals from 15 subjects under different video stimuli at a sampling rate of 1 kHz. Each subject participated in 3 sessions. In each session, each subject was required to watch 24 movie clips containing 4 different emotions: sad, neutral, fear, and happy. Each emotion contains a total of 6 movie clips, corresponding to 6 trials.

(3) \textbf{DEAP} \cite{koelstra2011deap} was constructed by Queen Mary University of London. The dataset has completely different acquisition device, emotional stimuli, data specifications, labeling approaches, experimental protocols, and subjects from the SEED-series datasets. Specifically, the DEAP dataset used a 128-channel Biosemi ActiveTwo System to record EEG signals from specific 32 channels of 32 subjects (16 males and 16 females) while watching 40 one-minute music videos at a sampling rate of 512 Hz. The 40 videos elicited different emotions according to the valence-arousal dimension. Specifically, the valence-arousal emotional model proposed by Russell \cite{russell1980circumplex} places each emotional state on a two-dimensional scale. The first dimension represents valence, ranging from negative to positive, and the second dimension represents arousal, ranging from calm to exciting. Participants rated valence and arousal using a continuous scale of 1 to 9 after watching each video clip.
}

\subsection{Pre-processing Procedures}{
For the SEED and SEED-IV datasets, the raw EEG signals were first downsampled to 200 Hz and filtered through a bandpass filter of 0.3-50 Hz to remove noise and artifacts. Then, the data were divided into multiple non-overlapping segments using a sliding window of 1s (SEED) and 4s (SEED-IV), respectively. After that, we extracted differential entropy (DE) features for each channel of each segment at five frequency bands: Delta (1-4 Hz), Theta (4-8 Hz), Alpha (8-14 Hz), Beta (14-31 Hz), and Gamma (31-50 Hz). Finally, the DE features from 62 channels and 5 bands were formed into a feature matrix of shape $62 \times 5$, which serves as input to the SCMM model. The extraction of DE features can be expressed as:
\begin{equation}
\label{eq: DE Features}
    DE (x) = \frac{1}{2} {\rm log}(2 \pi e \sigma^2),
\end{equation}
where $e$ is the Euler constant. $x$ is an EEG signal of a specific length that approximately obeys a Gaussian distribution $N(\mu, \sigma^2)$. Here, $\mu$ and $\sigma$ are the mean and standard deviation of $x$, respectively.

For the DEAP dataset, the raw EEG signals were initially downsampled to 128 Hz and denoised by a bandpass filter of 4-45 Hz. Subsequently, the data were segmented into multiple non-overlapping segments using a sliding window of 1s. Similar to the SEED and SEED-IV datasets, DE features were extracted for each channel of each segment at five frequency bands. Finally, the DE features from 32 channels and 5 bands were formed into a feature matrix of shape $32 \times 5$ as input to the model. In the experiments, we divided the continuous labels using a fixed threshold of 5 to convert them to binary classification tasks (low/high).}

\subsection{Handling Different Number of Channels}{
Since the SEED-series datasets and the DEAP dataset contain different numbers of electrodes (channels), we require channel processing before inputting DE features into the model. Specifically, we consider the fine-tuning dataset as the standard. When the number of channels in the fine-tuning dataset is less than in the pre-training dataset, we select data from the corresponding channels in the pre-training dataset and drop the data from the redundant channels as inputs (e.g., pre-training on SEED and fine-tuning on DEAP). Conversely, when the number of channels in the fine-tuning dataset is greater than in the pre-training dataset, we fill the missing channel data with zeros in the pre-training dataset to match the fine-tuning dataset (e.g., pre-training on DEAP and fine-tuning on SEED).}
}

\begin{table}[h]
    \caption{Hyperparameter settings of SCMM.}
    \centering
    \begin{tabular}{c|c|c}
        \toprule
        Hyperparameters & Pre-training & Fine-tuning \\ 
        \midrule
            Encoder & \multicolumn{2}{c}{3-layer 1D CNN} \\
        \midrule
            Projector & \multicolumn{2}{c}{2-layer MLP} \\
        \midrule
            Decoder & \multicolumn{2}{c}{single-layer MLP} \\
        \midrule
            Classifier & - & 2-layer MLP \\
        \midrule
            Masking Ratio $r$ & 0.5 & - \\
        \midrule
            Threshold $\mu$ & 0.1 & - \\
        \midrule
            Upper Bound $\alpha$ & 0.5 & - \\
        \midrule
            Sharpness $\tau_{\rm s}$ & 0.05 & - \\
        \midrule
            Temperature $\tau_{\rm c}$ & 0.5 & - \\
        \midrule
            Epoch & 200 & 50, 500 \\
        \midrule
            Optimizer & \multicolumn{2}{c}{Adam} \\
        \midrule
            Learning Rate & \multicolumn{2}{c}{$5 \times 10^{-4}$} \\
        \midrule
            Weight Decay & \multicolumn{2}{c}{$3 \times 10^{-4}$} \\
        \midrule
            Batch Size & 256 & 128 \\
        \bottomrule
    \end{tabular}
\label{tab:Hyperparameter Settings}
\end{table}

\begin{table*}[h]
    \caption{Experimental scenarios and data division for cross-corpus EEG-based emotion recognition.}
    \centering
    \begin{tabular}{c|c|c|c}
        \toprule
            Evaluations & Scenarios & Pre-training & Fine-tuning/Testing \\ 
        \midrule
            \multirow{2.5}{*}{Same-Class} & SEED-IV$^3$ $\rightarrow$ SEED$^3$ & SEED-IV, 3-class & SEED: 9/6 trials in each session of each subject \\
        \cmidrule(lr{0.01cm}){3-4}
            & SEED$^3$ $\rightarrow$ SEED-IV$^3$ & SEED & SEED-IV, 3-class: 12/6 trials in each session of each subject \\
        \midrule
            \multirow{6.5}{*}{Different-Class} & SEED-IV$^4$ $\rightarrow$ SEED$^3$ & SEED-IV, 4-class & SEED: 9/6 trials in each session of each subject \\
        \cmidrule(lr{0.01cm}){3-4}
            & SEED$^3$ $\rightarrow$ SEED-IV$^4$ & SEED & SEED-IV, 4-class: 16/8 trials in each session of each subject \\
        \cmidrule(lr{0.01cm}){2-4}
            & DEAP $\rightarrow$ SEED$^3$ & DEAP & SEED: 9/6 trials in each session of each subject \\
        \cmidrule(lr{0.01cm}){3-4}
            & SEED$^3$ $\rightarrow$ DEAP (Valence) & SEED & DEAP (Valence): 24/16 trials of each subject \\
        \cmidrule(lr{0.01cm}){3-4}
            & SEED$^3$ $\rightarrow$ DEAP (Arousal) & SEED & DEAP (Arousal): 24/16 trials of each subject \\
        \bottomrule
    \end{tabular}
\label{tab:Data Division}
\end{table*}

\section{Implementation Details}
\label{appendix:Implementation Details}{
To reduce computational load while maintaining model performance, we adopt a lightweight design for each module of SCMM. Specifically, we use a 3-layer 1D CNN for the encoder $E$ and a 2-layer MLP for the projector $P$. For the lightweight decoder $D$, we utilize a single-layer MLP for reconstruction. For the hyperparameter selection in the pre-training stage, we set $r$ to 0.5 and $\mu$ to 0.1 for hybrid masking, i.e., the ratio of random masking and channel masking is 9:1. We use the negative of cosine similarity as the metric function $Dist(\cdot, \cdot)$, and we set $\alpha$ to 0.5, $\tau_w$ to 0.05 and $\tau_c$ to 0.5 for soft contrastive learning. We use Adam optimizer with an initial learning rate of $5 \times 10^{-4}$ and an L2-norm penalty coefficient $3 \times 10^{-4}$. The pre-training process is conducted over 200 epochs with a batch size of 256. We save the model parameters $\theta$ from the final epoch as the pre-trained SCMM. In the fine-tuning stage, we input the encoded embeddings $h_i$ into an emotion classifier consisting of a 2-layer MLP for final emotion recognition. The Adam optimizer is utilized with an initial learning rate of $5 \times 10^{-4}$ and a weight decay of $3 \times 10^{-4}$. The number of fine-tuning epochs is set to 50 for SEED and SEED-IV and 500 for DEAP, with a batch size of 128. For efficient deployment and testing of the model, the pre-trained SCMM is optimized solely using cross-entropy loss during fine-tuning. All experiments are conducted using Python 3.9 with PyTorch 1.13 on an NVIDIA GeForce 3090 GPU. Table \ref{tab:Hyperparameter Settings} summarizes the hyperparameter settings of SCMM.
}

\section{Baselines and Experimental Setup}
\label{appendix:Experimental Settings}{
We compare the proposed SCMM against ten competitive baselines, including five transfer learning methods: DANN \cite{ganin2016domain}, BiDANN \cite{li2018novel}, TANN \cite{li2021novel}, PR-PL \cite{zhou2023pr}, and E$^2$STN \cite{zhou2025enhancing}, as well as five self-supervised learning models: SimCLR \cite{chen2020simple, tang2020exploring}, Mixup \cite{zhang2017mixup, wickstrom2022mixing}, TS-TCC \cite{eldele2021time}, MAE \cite{he2022masked}, and JCFA \cite{liu2024joint}. Note that E$^2$STN and JCFA are two state-of-the-art (SOTA) methods designed for cross-corpus EEG-based emotion recognition. To ensure a fair comparison, we adopt the same encoder, projector, decoder, and classifier structures as SCMM for SimCLR, Mixup, TS-TCC, and MAE. We use the default hyperparameters reported in the original papers for all models in the experiments, unless otherwise specified. Additionally, for DANN, BiDANN, TANN, PR-PL, E$^2$STN, MAE, and SCMM, the inputs are preprocessed 1-s DE features. In contrast, SimCLR, Mixup, TS-TCC, and JCFA use preprocessed 1-s EEG signals as inputs, in accordance with the specific design of each model.

In the experiments, we adopt a cross-corpus subject-independent protocol, consistent with the setup used by JCFA. Specifically, samples from one dataset are used for pre-training, while samples of each subject from another dataset are used individually for fine-tuning and testing. During the fine-tuning process, we use a leave-trials-out setting, where samples from a part of the trials in each session of each subject in the fine-tuning dataset are used for fine-tuning, and the remaining trials are used for testing. This approach effectively avoids information leakage. We calculate the average accuracy and standard deviation (ACC / STD $\%$) across all subjects in the test set to evaluate the model performance of SCMM. Table \ref{tab:Data Division} details the experimental settings for pre-training and fine-tuning.
}

\begin{figure*}[h]
    \begin{center}
        \includegraphics[width=1.0\textwidth]{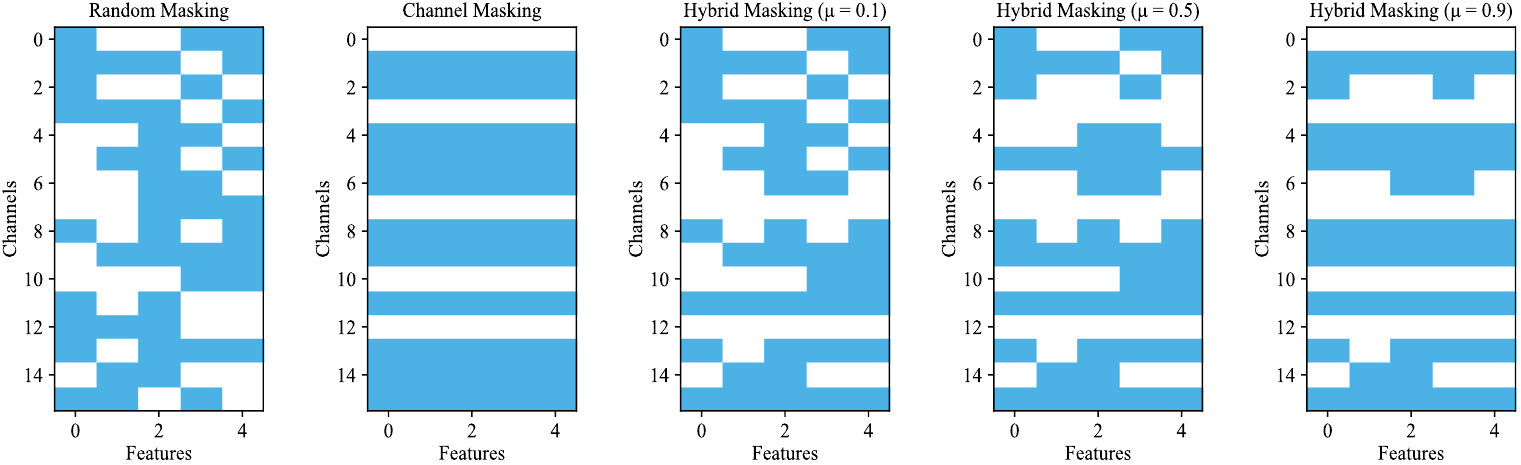}
    \end{center}
    \caption{Examples of generated masked samples using different masking strategies. The masking ratio is set to $r=0.5$, and the thresholds for hybrid masking are set to $\mu=0.1, 0.5$, and $0.9$, respectively. Best viewed in color.}
    \label{fig:Masking_Examples}
    \Description{Examples of generated masked samples using different masking strategies.}
\end{figure*}

\begin{table}[h]
    \caption{Comparison of different masking strategies on SEED and SEED-IV under the same-class condition.}
    \centering
    \begin{tabular}{c|c|c}
        \toprule
            Strategies & SEED-IV$^3$ $\rightarrow$ SEED$^3$ & SEED$^3$ $\rightarrow$ SEED-IV$^3$ \\ 
        \midrule
            Random & 90.30 / 08.80 & 84.63 / 10.99 \\
            Channel & 90.25 / 08.68 & 85.91 / 10.97 \\
            Parallel & 90.34 / 08.20 & 86.23 / 09.94 \\
            \textbf{Hybrid} & \textbf{91.61 / 07.56} & \textbf{87.24 / 08.35} \\
        \bottomrule
        \end{tabular}
\label{tab:Comparison of different masking strategies.}
\end{table}

\section{Masking Strategy}
\label{appendix:Masking Strategy}{
This paper introduces a novel hybrid masking strategy to generate diverse masked samples by considering both channel and feature relationships. To compare our approach with traditional masking strategies, we explore three different masking rules: random masking, channel masking, and hybrid masking. Figure \ref{fig:Masking_Examples} presents examples of
generated masked samples using three strategies.
\begin{itemize}
    \item \textbf{Random Masking:} Generate masks using a binomial distribution to randomly mask samples along the feature dimension, setting the values of masked features to 0.
    \item \textbf{Channel Masking:} Generate masks using a binomial distribution to randomly mask samples along the channel dimension, setting the values of all features within the masked channels to 0.
    \item \textbf{Hybrid Masking:} Generate a probability matrix using a uniform distribution that proportionally mixes masks generated by random masking and channel masking.
\end{itemize}

To assess the impact of various masking strategies, we conduct comparative experiments on the SEED and SEED-IV 3-category datasets using four strategies: random, channel, parallel, and hybrid masking. Specifically, the parallel masking strategy indicates that each sample is augmented randomly using one of the random masking or channel masking, and a threshold $\mu$ is used to control the probability of the two masking strategies being selected. Table \ref{tab:Comparison of different masking strategies.} presents the experimental results, showing that the hybrid masking strategy achieves the highest accuracy and lowest standard deviation in both experiments. This suggests that the integration of different masking approaches significantly improves the model performance and stability. In addition, parallel masking increases the richness of augmented samples by combining different strategies, which is slightly better than using a single masking approach. However, this strategy makes the model training process unstable, resulting in large standard deviations. In summary, our proposed hybrid masking strategy is highly flexible and can be extended to various datasets by integrating multiple masking strategies using different ratios, which is exceptionally suitable for data with rich semantic information. This strategy effectively generates more diverse masked samples, encouraging the model to comprehensively capture the inherent relationships of the data.
}

\begin{table*}[h]
    \caption{Subject-dependent and subject-independent EEG-based emotion recognition performance on SEED and SEED-IV.}
    \centering
    \begin{tabularx}{0.8\linewidth}{c| >{\centering\arraybackslash}X | >{\centering\arraybackslash}X | >{\centering\arraybackslash}X | >{\centering\arraybackslash}X}
        \toprule
        \multirow{2.5}{*}{Methods} & \multicolumn{2}{c|}{\textbf{Subject-Dependent}} & \multicolumn{2}{c}{\textbf{Subject-Independent}} \\
        \cmidrule(lr{0.01cm}){2-5}
        & SEED & SEED-IV & SEED & SEED-IV \\
        \midrule
            SVM \cite{suykens1999least} & 83.99 / 09.72  & 56.61 / 20.02 & 56.73 / 16.29 & 37.99 / 12.52 \\
            TCA \cite{pan2010domain} & -  & - & 63.38 / 14.88 & 37.01 / 10.47 \\
            GFK \cite{gong2012geodesic} & - & - & 71.31 / 14.09 & 44.04 / 09.31 \\
            M3D \cite{luo2025m3d} & - & - & 84.57 / 09.49 & 60.94 / 08.84 \\
        \midrule
             \multicolumn{5}{c}{\textbf{Deep Learning}} \\ 
        \midrule
            DBN \cite{zheng2015investigating} & 86.08 / 08.34 & 66.77 / 07.38 & - & - \\
            DANN \cite{ganin2016domain} & 83.99 / 09.72 & 56.73 / 16.29 & 56.61 / 20.02 & 37.99 / 12.52 \\
            DAN \cite{li2018cross} & - & - & 83.81 / 08.56 & 58.87 / 08.13 \\
            DGCNN \cite{song2018eeg} & 90.40 / 08.49 & 69.88 / 16.29 & 79.95 / 09.02 & 52.82 / 09.23 \\
            BiDANN \cite{li2018novel} & 92.38 / 07.04 & 70.29 / 12.63 & 84.14 / 06.87 & 65.59 / 10.39 \\
            BiHDM \cite{li2020novel} & 93.12 / 06.06 & 74.35 / 14.09 & 85.40 / 07.53 & 69.03 / 08.66 \\
            RGNN \cite{zhong2020eeg} & 94.24 / 05.95 & 79.37 / 10.54 & 85.30 / 06.72 & 73.84 / 08.02 \\
            PR-PL \cite{zhou2023pr} & 94.84 / 09.16 & 83.33 / 10.61 & 93.06 / 05.12 & 81.32 / 08.53 \\
            PGCN \cite{jin2024pgcn} & 96.93 / 05.11 & 82.24 / 14.85 & 84.59 / 08.68 & 73.69 / 07.16 \\
        \midrule
             \multicolumn{5}{c}{\textbf{Self-Supervised Learning}} \\ 
        \midrule
            SimCLR \cite{chen2020simple} & 81.79 / 11.15 & 52.47 / 11.57 & 63.45 / 15.96 & 50.07 / 11.17 \\
            MoCo \cite{he2020momentum} & 76.58 / 10.72 & 49.40 / 10.99 & 58.26 / 15.05 & 46.19 / 10.04 \\
            SSL-EEG \cite{xie2021novel} & 83.32 / 09.20 & 63.59 / 19.82 & 67.52 / 12.73 & 53.62 / 08.47 \\
            GMSS \cite{li2022gmss} & 89.18 / 09.74 & 65.61 / 17.33 & 76.04 / 11.91 & 62.13 / 08.33 \\
        \midrule
            \textbf{SCMM (P-T)} & \textbf{92.19 / 07.68} & \textbf{81.41 / 08.17} & \textbf{83.84 / 06.22} & \textbf{75.28 / 06.37} \\
            \textbf{SCMM (E-E)} & \textbf{93.02 / 06.67} & \textbf{81.60 / 08.12} & \textbf{84.88 / 05.85} & \textbf{76.09 / 06.77} \\
        \bottomrule
    \end{tabularx}
\label{tab:Model performance for subject-dependent and subject-independent EEG-based emotion recognition on SEED and SEED-IV.}
\end{table*}

\section{Within-Dataset Validation}
\label{appendix:Within-Dataset Validation}{
To evaluate the model performance of SCMM for EEG-based emotion recognition within a single dataset, we conduct additional experiments on SEED and SEED-IV based on two experimental protocols: subject-dependent and subject-independent. In the experiments, we compare the proposed SCMM with three different types of models: machine learning, deep learning, and self-supervised learning. Additionally, we explore the model performance of SCMM using two different training strategies: pre-training-fine-tuning (P-T) and end-to-end (E-E). Table \ref{tab:Model performance for subject-dependent and subject-independent EEG-based emotion recognition on SEED and SEED-IV.} presents the experimental results, indicating that our model achieves competitive results compared with various advanced deep learning models and significantly outperforms traditional machine learning methods and self-supervised learning models. In addition, the SCMM model using the end-to-end training strategy is better than the pre-training-fine-tuning strategy due to the introduction of the emotion classifier for joint learning. In summary, comprehensive experimental results demonstrate the effectiveness of the proposed SCMM in within-dataset EEG-based emotion recognition tasks.
}

\section{Full Results}
\label{appendix:Full Results}{
Tables \ref{tab:Full results on SEED, SEED-IV and DEAP under two cross-corpus conditions: same-class and different-class.} to \ref{tab:Full results of hyperparameter sensitivity analysis on SEED and SEED-IV.} present the full experimental results of SCMM.
\begin{table*}[h]
    \caption{Full results on SEED, SEED-IV, and DEAP under same-class and different-class cross-corpus conditions.}
    \centering
    \begin{tabularx}{\linewidth}{c| >{\centering\arraybackslash}X | >{\centering\arraybackslash}X | >{\centering\arraybackslash}X | >{\centering\arraybackslash}X | >{\centering\arraybackslash}X | >{\centering\arraybackslash}X}
        \toprule
        Scenarios & Accuracy & Precision & Recall & F1 Score & AUROC & AUPRC \\ 
        \midrule
            \multicolumn{7}{c}{\textbf{Same-Class}} \\
        \midrule
            SEED-IV$^3$ $\rightarrow$ SEED$^3$ & 91.61 / 07.56 & 93.38 / 05.71 & 91.35 / 07.89 & 91.02 / 08.40 & 95.45 / 05.82 & 91.90 / 09.71 \\
            SEED$^3$ $\rightarrow$ SEED-IV$^3$ & 87.24 / 08.35 & 87.10 / 09.02 & 85.58 / 09.59 & 84.39 / 10.07 & 89.91 / 07.78 & 86.81 / 10.12 \\
        \midrule
            \multicolumn{7}{c}{\textbf{Different-Class}} \\
        \midrule
            SEED-IV$^4$ $\rightarrow$ SEED$^3$ & 91.26 / 07.91 & 92.62 / 06.79 & 91.02 / 08.14 & 90.77 / 08.69 & 95.59 / 06.12 & 92.28 / 09.46 \\
            SEED$^3$ $\rightarrow$ SEED-IV$^4$ & 80.89 / 08.69 & 78.02 / 08.39 & 82.84 / 06.68 & 77.08 / 07.72 & 89.98 / 06.09 & 83.96 / 08.68 \\
            DEAP$^3$ $\rightarrow$ SEED$^3$ & 91.70 / 08.07 & 92.94 / 06.85 & 91.48 / 08.29 & 91.15 / 08.88 & 95.45 / 06.64 & 92.34 / 10.49 \\
            SEED$^3$ $\rightarrow$ DEAP (Valence) & 73.96 / 06.75 & 68.11 / 08.89 & 65.77 / 07.75 & 65.68 / 08.75 & 74.15 / 07.34 & 72.12 / 07.68 \\
            SEED$^3$ $\rightarrow$ DEAP (Arousal) & 72.66 / 05.67 & 70.97 / 05.91 & 68.90 / 07.11 & 68.84 / 07.57 & 75.77 / 07.33 & 74.79 / 07.50 \\
    \bottomrule
    \end{tabularx}
\label{tab:Full results on SEED, SEED-IV and DEAP under two cross-corpus conditions: same-class and different-class.}
\end{table*}

\begin{table*}[h]
    \caption{Full results of ablation study on SEED and SEED-IV under same-class and different-class cross-corpus conditions.}
    \centering
    \begin{tabularx}{\linewidth}{c| >{\centering\arraybackslash}X | >{\centering\arraybackslash}X | >{\centering\arraybackslash}X | >{\centering\arraybackslash}X | >{\centering\arraybackslash}X | >{\centering\arraybackslash}X}
        \toprule
        Scenarios & Accuracy & Precision & Recall & F1 Score & AUROC & AUPRC \\ 
        \midrule
           \multicolumn{7}{c}{\textbf{w/o Soft Contrastive Loss $\mathcal{L}_{\rm C}$}} \\
        \midrule
            SEED-IV$^3$ $\rightarrow$ SEED$^3$ & 89.68 / 09.32 & 91.81 / 07.76 & 89.39 / 09.58 & 89.20 / 09.97 & 94.71 / 05.95 & 91.73 / 09.32 \\
            SEED$^3$ $\rightarrow$ SEED-IV$^3$ & 84.24 / 11.90 & 84.79 / 14.99 & 83.84 / 10.79 & 81.94 / 14.11 & 90.65 / 10.10 & 85.69 / 15.93 \\
            SEED-IV$^4$ $\rightarrow$ SEED$^3$ & 89.45 / 09.10 & 91.23 / 08.34 & 89.21 / 09.32 & 89.13 / 09.58 & 95.11 / 05.86 & 91.24 / 10.77 \\
            SEED$^3$ $\rightarrow$ SEED-IV$^4$ & 77.24 / 09.14 & 73.21 / 14.36 & 73.55 / 12.89 & 70.24 / 14.29 & 88.57 / 09.82 & 80.79 / 16.07 \\
        \midrule
           \multicolumn{7}{c}{\textbf{w/o Reconstruction Loss $\mathcal{L}_{\rm R}$}} \\
        \midrule
            SEED-IV$^3$ $\rightarrow$ SEED$^3$ & 90.73 / 08.48 & 93.22 / 06.17 & 90.46 / 08.76 & 90.26 / 09.07 & 94.57 / 06.58 & 90.52 / 10.06 \\
            SEED$^3$ $\rightarrow$ SEED-IV$^3$ & 85.07 / 11.05 & 85.26 / 14.01 & 81.59 / 13.58 & 80.88 / 15.46 & 88.35 / 15.99 & 85.16 / 18.10 \\
            SEED-IV$^4$ $\rightarrow$ SEED$^3$ & 90.96 / 08.36 & 92.43 / 07.24 & 90.70 / 08.57 & 90.41 / 09.07 & 94.36 / 06.39 & 89.98 / 10.40 \\
            SEED$^3$ $\rightarrow$ SEED-IV$^4$ & 78.32 / 07.19 & 73.07 / 14.36 & 74.59 / 12.07 & 72.01 / 13.92 & 87.02 / 12.49 & 82.09 / 16.17 \\
        \midrule
           \multicolumn{7}{c}{\textbf{Full Model}} \\
        \midrule
            SEED-IV$^3$ $\rightarrow$ SEED$^3$ & 91.61 / 07.56 & 93.38 / 05.71 & 91.35 / 07.89 & 91.02 / 08.40 & 95.45 / 05.82 & 91.90 / 09.71 \\
            SEED$^3$ $\rightarrow$ SEED-IV$^3$ & 87.24 / 08.35 & 87.10 / 09.02 & 85.58 / 09.59 & 84.39 / 10.07 & 89.91 / 07.78 & 86.81 / 10.12 \\
            SEED-IV$^4$ $\rightarrow$ SEED$^3$ & 91.26 / 07.91 & 92.62 / 06.79 & 91.02 / 08.14 & 90.77 / 08.69 & 95.59 / 06.12 & 92.28 / 09.46 \\
            SEED$^3$ $\rightarrow$ SEED-IV$^4$ & 80.89 / 08.69 & 78.02 / 08.39 & 82.84 / 06.68 & 77.08 / 07.72 & 89.98 / 06.09 & 83.96 / 08.68 \\
        \bottomrule
    \end{tabularx}
\label{tab:Full results of ablation study on SEED and SEED-IV under same-class and different-class cross-corpus conditions.}
\end{table*}

\begin{table*}[h]
    \caption{Full results of model performance with limited labeled data for fine-tuning on SEED and SEED-IV under same-class and different-class cross-corpus conditions.}
    \centering
    \begin{tabularx}{\linewidth}{c| >{\centering\arraybackslash}X | >{\centering\arraybackslash}X | >{\centering\arraybackslash}X | >{\centering\arraybackslash}X | >{\centering\arraybackslash}X | >{\centering\arraybackslash}X}
        \toprule
        Scenarios & Accuracy & Precision & Recall & F1 Score & AUROC & AUPRC \\ 
        \midrule
           \multicolumn{7}{c}{\textbf{1\% of Labeled Data}} \\
        \midrule
            SEED-IV$^3$ $\rightarrow$ SEED$^3$ & 74.98 / 17.00 & 74.52 / 21.19 & 74.69 / 17.11 & 70.64 / 21.53 & 93.76 / 06.20 & 89.32 / 09.68 \\
            SEED-IV$^4$ $\rightarrow$ SEED$^3$ & 73.20 / 16.09 & 73.41 / 21.28 & 72.93 / 16.23 & 69.14 / 20.07 & 94.32 / 04.69 & 90.31 / 07.53 \\
            SEED$^3$ $\rightarrow$ SEED-IV$^3$ & 71.35 / 16.29 & 72.22 / 18.40 & 69.37 / 15.47 & 66.23 / 18.19 & 80.18 / 14.09 & 74.07 / 17.64 \\
            SEED$^3$ $\rightarrow$ SEED-IV$^4$ & 60.59 / 23.72 & 55.23 / 29.74 & 58.65 / 22.49 & 52.48 / 27.81 & 82.19 / 13.44 & 70.54 / 19.49 \\
        \midrule
           \multicolumn{7}{c}{\textbf{5\% of Labeled Data}} \\
        \midrule
            SEED-IV$^3$ $\rightarrow$ SEED$^3$ & 95.23 / 06.98 & 96.20 / 04.72 & 95.10 / 07.14 & 94.86 / 07.88 & 99.67 / 00.38 & 99.45 / 00.62 \\
            SEED-IV$^4$ $\rightarrow$ SEED$^3$ & 90.51 / 12.43 & 92.90 / 08.38 & 90.27 / 12.70 & 89.04 / 15.29 & 98.23 / 02.25 & 96.76 / 04.34 \\
            SEED$^3$ $\rightarrow$ SEED-IV$^3$ & 87.20 / 16.12 & 88.31 / 14.88 & 86.08 / 15.69 & 85.35 / 17.64 & 91.93 / 10.70 & 89.14 / 13.71 \\
            SEED$^3$ $\rightarrow$ SEED-IV$^4$ & 86.01 / 09.76 & 84.86 / 14.77 & 82.69 / 11.35 & 81.29 / 13.75 & 95.14 / 03.37 & 92.56 / 06.89 \\
        \midrule
           \multicolumn{7}{c}{\textbf{10\% of Labeled Data}} \\
        \midrule
            SEED-IV$^3$ $\rightarrow$ SEED$^3$ & 99.86 / 00.31 & 99.87 / 00.30 & 99.86 / 00.31 & 99.86 / 00.31 & 100.00 / 00.00 & 100.00 / 00.00  \\
            SEED-IV$^4$ $\rightarrow$ SEED$^3$ & 99.72 / 00.80 & 99.74 / 00.74 & 99.71 / 00.81 & 99.72 / 00.80 & 99.87 / 00.46 & 99.78 / 00.77 \\
            SEED$^3$ $\rightarrow$ SEED-IV$^3$ & 90.04 / 16.94 & 91.00 / 17.05 & 90.09 / 15.87 & 89.32 / 18.21 & 94.29 / 10.86 & 92.55 / 13.98 \\
            SEED$^3$ $\rightarrow$ SEED-IV$^4$ & 94.43 / 05.60 & 95.66 / 04.71 & 92.18 / 08.23 & 92.19 / 08.82 & 98.12 / 02.13 & 96.03 / 05.45 \\
        \midrule
           \multicolumn{7}{c}{\textbf{20\% of Labeled Data}} \\
        \midrule
            SEED-IV$^3$ $\rightarrow$ SEED$^3$ & 100.00 / 00.00 & 100.00 / 00.00 & 100.00 / 00.00 & 100.00 / 00.00 & 100.00 / 00.00 & 100.00 / 00.00 \\
            SEED-IV$^4$ $\rightarrow$ SEED$^3$ & 100.00 / 00.00 & 100.00 / 00.00 & 100.00 / 00.00 & 100.00 / 00.00 & 100.00 / 00.00 & 100.00 / 00.00 \\
            SEED$^3$ $\rightarrow$ SEED-IV$^3$ & 94.16 / 08.52 & 95.20 / 06.36 & 93.66 / 08.00 & 93.72 / 08.35 & 97.59 / 03.38 & 96.53 / 04.85 \\
            SEED$^3$ $\rightarrow$ SEED-IV$^4$ & 98.91 / 02.62 & 98.82 / 02.97 & 98.41 / 03.73 & 98.52 / 03.65 & 98.97 / 03.20 & 98.13 / 05.64 \\
        \bottomrule
    \end{tabularx}
\label{tab:Full results of model performance with limited labeled data for fine-tuning on SEED and SEED-IV.}
\end{table*}

\begin{table*}[h]
    \caption{Full results of hyperparameter sensitivity analysis on SEED and SEED-IV under the same-class cross-corpus condition.}
    \centering
    \begin{tabular}{c|c|c}
        \toprule
            Hyperparameters & SEED-IV$^3$ $\rightarrow$ SEED$^3$ & SEED$^3$ $\rightarrow$ SEED-IV$^3$ \\
        \midrule
           \multicolumn{3}{c}{\textbf{Masking Ratio $r$}} \\
        \midrule
            0.1 & 91.19 / 08.07 & 85.05 / 11.68 \\
            0.25 & 91.50 / 07.69 & 85.28 / 09.91 \\
            \textbf{0.5} & \textbf{91.61 / 07.56} & \textbf{87.24 / 08.35} \\
            0.75 & 90.24 / 08.08 & 83.49 / 13.19 \\
        \midrule
           \multicolumn{3}{c}{\textbf{Threshold $\mu$}} \\
        \midrule
            0 (Random) & 90.30 / 08.80 & 84.63 / 10.99 \\
            \textbf{0.1} & \textbf{91.61 / 07.56} & \textbf{87.24 / 08.35} \\
            0.2 & 90.93 / 08.18 & 86.51 / 08.44 \\
            0.3 & 90.84 / 07.70 & 86.66 / 09.98 \\
            0.4 & 89.34 / 08.70 & 86.94 / 09.81 \\
            0.5 & 91.00 / 08.22 & 87.16 / 11.09 \\
            0.6 & 89.93 / 08.90 & 85.83 / 09.65 \\
            0.7 & 91.08 / 07.64 & 85.14 / 11.60 \\
            0.8 & 89.52 / 08.29 & 86.10 / 12.48 \\
            0.9 & 90.25 / 08.68 & 85.91 / 10.97 \\
            1 (Channel) & 90.25 / 08.68 & 85.91 / 10.97 \\
        \midrule
           \multicolumn{3}{c}{\textbf{Metric Function $Dist(\cdot, \cdot)$}} \\
        \midrule
            Manhattan & 90.09 / 09.06 & 85.83 / 11.14 \\
            Euclidean & 90.86 / 08.57 & 85.40 / 13.51 \\
            \textbf{Cosine} & \textbf{91.61 / 07.56} & \textbf{87.24 / 08.35} \\
        \midrule
           \multicolumn{3}{c}{\textbf{Upper Bound $\alpha$}} \\
        \midrule
            0.25 & 91.21 / 08.54 & - \\
            \textbf{0.5} & \textbf{91.61 / 07.56} & - \\
            0.75 & 91.12 / 08.25 & - \\
            1 & 90.50 / 08.00 & - \\
        \midrule
           \multicolumn{3}{c}{\textbf{Sharpness $\tau_{\rm s}$}} \\
        \midrule
            0.01 & 90.93 / 09.13 & - \\
            \textbf{0.05} & \textbf{91.61 / 07.56} & - \\
            0.1 & 91.08 / 09.00 & - \\
            0.2 & 90.59 / 07.40 & - \\
            0.5 & 90.63 / 09.16 & - \\
            1 & 90.09 / 08.62 & - \\
            2 & 89.58 / 09.26 & - \\
        \midrule
           \multicolumn{3}{c}{\textbf{Temperature $\tau_{\rm c}$}} \\
        \midrule
            0.05 & 90.97 / 08.17 & 85.95 / 11.14 \\
            0.2 & 91.16 / 07.85 & 86.82 / 11.46 \\
            \textbf{0.5} & \textbf{91.61 / 07.56} & \textbf{87.24 / 08.35} \\
            2 & 90.62 / 08.66 & 86.73 / 10.18 \\
        \midrule
            \multicolumn{3}{c}{\textbf{Pre-training Batch Size}} \\
        \midrule
            8 & 91.41 / 08.42 & 86.64 / 11.77 \\
            16 & 91.29 / 08.90 & 86.33 / 12.04 \\
            32 & 91.18 / 08.23 & 86.77 / 12.72 \\
            64 & 90.35 / 08.37 & 86.85 / 10.96 \\
            128 & 91.00 / 08.22 & 85.74 / 13.42 \\
            \textbf{256} & \textbf{91.61 / 07.56} & \textbf{87.24 / 08.35} \\
            512 & 90.04 / 08.04 & 86.74 / 11.14 \\
        \bottomrule
    \end{tabular}
\label{tab:Full results of hyperparameter sensitivity analysis on SEED and SEED-IV.}
\end{table*}
}

\end{document}